\documentclass[aps,twocolumn,secnumarabic,amsmath,amssymb,superscriptaddress,floatfix,eqsecnum,graphicx,dvipdfmx]{revtex4-2}
\usepackage{amsthm}
\usepackage{amsmath}
\usepackage{amsfonts}
\usepackage{amssymb}
\usepackage{mathtools} % to use \coloneqq
\usepackage{bm} % bold math
\usepackage{flafter}
\usepackage{slashed} 
\usepackage{color}
\usepackage{bm}
\usepackage{bbm}
\usepackage[breaklinks=true,colorlinks,citecolor=blue,linkcolor=blue,urlcolor=blue]{hyperref}
\usepackage{cases}
\usepackage{float}
\usepackage{url}
\usepackage{framed}
\usepackage{cancel,soul,ulem}
\usepackage{appendix}
\usepackage{graphicx}
\usepackage{multirow}
\usepackage{tikz}
\tikzset{>=latex} % for LaTeX arrow head
\usetikzlibrary{arrows.meta}
\usepackage{pgfplots} % for the axis environment
\pgfplotsset{compat = newest} % for latest version

\def\:={\,\raisebox{0.85pt}{.}\hspace{-2.78pt}\raisebox{2.85pt}{.}\!\!=\,}
\def\=:{\,=\!\!\raisebox{0.85pt}{.}\hspace{-2.78pt}\raisebox{2.85pt}{.}\,}

\usepackage{ulem}

\begin{document}
\title{
Correlation functions at the topological quantum phase transition in the $S=1$ XXZ chain with single-ion anisotropy
}

\author{Toshiya Hikihara}
\affiliation{Graduate School of Science and Technology, Gunma University, Kiryu, Gunma 376-8515, Japan}
\author{Akira Furusaki}
\affiliation{RIKEN Center for Emergent Matter Science, Wako, Saitama, 351-0198, Japan}
%\affiliation{Condensed Matter Theory Laboratory, RIKEN, Wako, Saitama, 351-0198, Japan}

\begin{abstract}
We study the one-dimensional $S=1$ XXZ spin model with single-ion anisotropy.
It is known that at the transition points between the Haldane and large-$D$ phases,
the model exhibits a quantum criticality described by the Gaussian theory, i.e., a conformal field theory with the central charge $c=1$.
Using the bosonization approach, we investigate various correlation functions at the phase transition and derive their asymptotic forms.
This allows us to clarify their peculiar behavior: the longitudinal (transverse) two-point spin correlation function has components that decay algebraically only in the uniform (staggered) sector.
These theoretical predictions are verified by the numerical calculations using the density-matrix renormalization group method.
The effect of weak bond alternation on the critical ground state at the phase transition is also discussed.
It is shown that the bond alternation induces the missing power-law components in the correlation functions. 
\end{abstract}

\date{\today}

\maketitle

\section{Introduction}\label{sec:Intro}

One-dimensional (1D) quantum magnets have been studied extensively for many decades since they exhibit a variety of exotic physical phenomena induced by strong quantum fluctuations.
One of the prominent examples is the Haldane phase \cite{Haldane1983a,Haldane1983b}, which has a quantum disordered ground state with an energy gap and is characterized by a non-local string order parameter \cite{dNijsR1989} and a hidden spontaneous $Z_2 \times Z_2$ symmetry breaking \cite{KennedyT1992}.
In recent years, the Haldane phase has regained much attention as a symmetry-protected topological phase \cite{PollmannBTAO2012}.
Another notable example is a quantum critical state realized at a continuous phase transition between different phases.
Due to strong quantum critical fluctuations, the quantum critical state exhibits peculiar behaviors such as diverging correlation lengths and algebraically decaying correlation functions \cite{GogolinNT1998,Giamarchi2004,SchulzCP1998,Affleck1990}.
The quantum critical points between phases that spontaneously break different symmetries have also attracted much interest
in the context of deconfined quantum criticality \cite{JiangM2019,MudryFMH2019}.
%{\bf (**** Appropriate refs. for DQC in 1D? Or, should we remove this sentence?****)}

In this paper, we study the $S=1$ XXZ spin chain with single-ion anisotropy.
The model Hamiltonian is given by
\begin{equation}
H=\sum_{j\in \mathbb{Z}}\left[
S^x_jS^x_{j+1}+S^y_jS^y_{j+1}+\Delta S^z_jS^z_{j+1}
+D(S^z_j)^2
\right],
\label{eq:Ham}
\end{equation}
where $\bm{S}_j=(S^x_j,S^y_j,S^z_j)$ is the $S=1$ spin operator on the $j$th site.
We have set the overall energy scale, the exchange coupling, $J=1$, for simplicity.
%the SU(2) spin operator at $j$th site obeying the algebra
%\begin{equation}
%[S^\alpha_j, S^\beta_k]=i\delta_{j,k}\epsilon^{\alpha\beta\gamma}S^\gamma_j,
%\qquad
%\bm{S}_j\cdot\bm{S}_j=2.
%\end{equation}
%We also define $S_j^\pm=S_j^x \pm i S_j^y$.
%\begin{equation}
%S^+_j=S^x_j + i S^y_j,
%\qquad
%S^-_j=S^x_j - i S^y_j.
%\end{equation}
%The Hamiltonian (\ref{eq:Ham}) has the global U(1) symmetry generated by
%\begin{equation}
%U_z=\exp\!\left(it\sum_jS^z_j\right),
%\end{equation}
%where $t\in\mathbb{R}$.
The ground-state phase diagram of the model (\ref{eq:Ham}) has a rich structure including the Haldane phase, large-$D$ phase, N\'{e}el phase, Ferromagnetic phase, and two types (XY1 and XY2) of Tomonaga-Luttinger liquid (TLL) phases in the two-dimensional parameter space $(\Delta, D)$ \cite{Schulz1986,dNijsR1989,ChenHS2003}.
In particular, it has been shown \cite{Schulz1986,ChenHS2003} that the model exhibits the Gaussian quantum criticality along the boundary between the Haldane and large-$D$ phases.
This phase transition is of considerable interest as it represents a transition between a symmetry-protected topological phase and a topologically trivial phase.
Nevertheless, to the best of our knowledge, a comprehensive analysis of the critical behavior of correlation functions has not been reported, except for the study of the model with additional bond alternation \cite{EjimaYELOF2018}. This analysis is of fundamental importance for understanding the nature of the transition.

In this paper, we revisit the transition between the Haldane and large-$D$ phases and calculate several correlation functions at criticality, using the bosonization method \cite{Schulz1986} and density-matrix renormalization group (DMRG) method \cite{White1992,White1993}.
First, we show using bosonization that the correlation functions exhibit the following characteristic behavior.
The correlation functions of the longitudinal spin, dimer, and squared-spin operators exhibit the algebraically decaying behavior only in their uniform components, while their staggered components decay exponentially.
In contrast, the transverse-spin correlation function contains the algebraically decaying terms only in the staggered components, with the uniform components decaying exponentially.
These behaviors of the correlation functions are remarkably distinct from those in conventional critical antiferromagnets, such as the $S=1/2$ antiferromagnetic XXZ spin chain.
Then, to confirm these analytic results, we perform the numerical calculation of the correlation functions using the DMRG method.
By fitting the numerical data to the analytic forms, we demonstrate the validity of the effective theory and determine quantities such as the TLL parameter.
We also consider the effect of weak bond alternation on the critical theory and show that the aforementioned exponentially decaying components in the correlation functions become algebraically decaying in the presence of the bond alternation, in a manner similar to that observed in the $S=1/2$ XXZ spin chain.

The rest of the paper is organized as follows.
In Sec.\ \ref{sec:effective_theory}, we review the effective theory for the quantum criticality at the transition between the Haldane and large-$D$ phases and derive analytic forms of correlation functions.
The numerical results of the DMRG calculation are presented and discussed in Sec.\ \ref{sec:Num}.
The paper is concluded in Sec.\ \ref{sec:conc}.
Appendices\ \ref{app:bosonization_details} and \ref{app:finite_open_chain} present some details of bosonization calculations to derive the effective theory and correlation functions.
The numerical procedure for determining the critical points between the Haldane and large-$D$ phases is described in Appendix\ \ref{app:crt_point}.

\section{Low-energy Effective theory}\label{sec:effective_theory}
\subsection{Bosonization}\label{subsec:bosonization}

In this section, in order to set the stage and fix the notations, we review the low-energy effective theory for the quantum criticality at the transition between the Haldane and large-$D$ phases in the $S=1$ spin model (\ref{eq:Ham}), following the bosonization approach introduced by Schulz \cite{Schulz1986}.
We will see that the effective theory is the Gaussian model, i.e., a conformal field theory (CFT) with the central charge $c=1$\cite{FrancescoMS1997,Ginsparg1988}.
Some details of the calculations are presented in Appendix \ref{app:bosonization_details}.

In the Schulz's bosonization approach, the $S=1$ spin chain is represented by two $S=\frac12$ spin chains.
The spin operator $\bm{S}_j$ on the $j$th site is written in terms of two spin-$\frac12$ operators $\bm{s}_{n,j}=(s_{n,j}^x,s_{n,j}^y,s_{n,j}^z)$, $n=1,2$ as
% with $n=1,2$:
\begin{equation}
S_j^\alpha=s^\alpha_{1,j}+s^\alpha_{2,j}.
\label{S=s1+s2}
\end{equation}
The two independent spin-$\frac{1}{2}$ chains can be bosonized in a standard fashion.
Here, we use the notation of Refs.~\cite{HikiharaF2001,MudryFMH2019}
and write the spin operators as
\begin{subequations}
\label{eq:s_n,j}
\begin{align}
s^z_{n,j}&=
\frac{1}{\sqrt\pi}\frac{d\phi_n(x)}{dx}+(-1)^j a\sin[\sqrt{4\pi}\phi_n(x)],
\label{s^z_n,j}
\\
s^+_{n,j}&=
\exp[i\sqrt{\pi}\theta_n(j)]
\left[b(-1)^j+\tilde{b}\sin[\sqrt{4\pi}\phi_n(j)]\right],
\label{s^+_n,j}
\\
s^-_{n,j}&=
\left[b(-1)^j+\tilde{b}\sin[\sqrt{4\pi}\phi_n(j)]\right]\exp[-i\sqrt{\pi}\theta_n(j)]
\nonumber\\
&=
\exp[-i\sqrt{\pi}\theta_n(j)]\left[b(-1)^j-\tilde{b}\sin[\sqrt{4\pi}\phi_n(j)]\right],
\label{s^-_n,j}
\end{align}
\end{subequations}
where $a$, $b$, and $\tilde{b}$ are real numbers.
The bosonic fields $\phi_n(x)$ and $\theta_n(x)$ are defined on the one-dimensional space $x\in\mathbb{R}$ and
obey the commutation relations
%\begin{equation}
$[\phi_n(x),\theta_{n'}(y)]=-\frac{i}{2}\delta_{n,n'}[1+\mathrm{sgn}(x-y)].$
%\end{equation}
The fields $\phi_n(x)$ are compactified such that
$\phi_n(x)\equiv\phi_n(x)+2\pi R$ with $R=1/\sqrt{4\pi}$.

Using the bosonic-field representation of the spin operators, Eq.\ (\ref{eq:s_n,j}), and introducing the linear combinations of the bosonic fields,
\begin{subequations}
\label{eq:phi_theta_pm}
\begin{align}
&\phi_\pm(x)=\frac{1}{\sqrt2}\left[\phi_1(x)\pm\phi_2(x)\right],
%\qquad
\\
&\theta_\pm(x)=\frac{1}{\sqrt2}\left[\theta_1(x)\pm\theta_2(x)\right],
\end{align}
\end{subequations}
we obtain the bosonized effective Hamiltonian for the $S=1$ spin chain (\ref{eq:Ham}) as 
\begin{equation}
H=H_++H_-+H_\pm,
\label{eq:sum_H+_H-_Hpm}
\end{equation}
where
\begin{subequations}
\label{eq:boson_H+_H-_Hpm}
\begin{align}
H_+=&
\int dx\,\biggl\{
\frac{v_+}{2}\biggl[\frac{1}{K_+}\left(\frac{d\phi_+}{dx}\right)^2
+K_+\left(\frac{d\theta_+}{dx}\right)^2\biggr]
\nonumber \\
&\qquad\quad+a^2(\Delta-D)\cos(\sqrt{8\pi}\phi_+)
\biggl\},
\label{H_+ with K_+}
\\
H_-=&
\int dx\,\biggl\{
\frac{v_-}{2}\biggl[\frac{1}{K_-}\left(\frac{d\phi_-}{dx}\right)^2
+K_-\left(\frac{d\theta_-}{dx}\right)^2\biggr]
\nonumber \\
&\qquad\quad
-a^2(\Delta-D)\cos(\sqrt{8\pi}\phi_-)
\nonumber \\
&\qquad\quad
-2b^2\cos(\sqrt{2\pi}\theta_-)
\nonumber\\
&\qquad\quad
-\tilde{b}^2\cos(\sqrt{2\pi}\theta_-)\cos(\sqrt{8\pi}\phi_-)
\biggr\},
\label{H_- with K_-}
\\
H_\pm=&
\int dx\,\biggl[
a^2 \Delta \cos(\sqrt{8\pi}\phi_+)\cos(\sqrt{8\pi}\phi_-)
\nonumber \\
&\qquad\quad
+\tilde{b}^2\cos(\sqrt{2\pi}\theta_-)\cos(\sqrt{8\pi}\phi_+)
\biggl].
\label{H_pm}
\end{align}
\end{subequations}
Here, the TLL parameters $K_\pm$ and the velocities $v_\pm$ are defined by
\begin{subequations}
\begin{align}
%&
%v_+=\left(1+\frac{2}{\pi}(3\Delta+D)\right)^{1/2},
%\label{v_+}\\
%&
%v_-=\left(1+\frac{2}{\pi}(\Delta-D)\right)^{1/2},
%\label{v_-}\\
&
K_+=\left[1+\frac{2}{\pi}(3\Delta+D)\right]^{-1/2},~~~v_+=\frac{1}{K_+},
\label{K_+andv_+}\\
&
K_-=\left[1+\frac{2}{\pi}(\Delta-D)\right]^{-1/2},~~~v_-=\frac{1}{K_-}.
\label{K_-andv_-}
\end{align}
\end{subequations}
The effective Hamiltonian $H_++H_-+H_\pm$ is equivalent to the one
derived by Schulz \cite{Schulz1986}.

The Haldane-large $D$ phase transition line of our main interest is located in the first quadrant,
$\Delta>0$ and $D>0$, according to the phase diagarm in Ref.\ \cite{ChenHS2003}.
Following Schulz \cite{Schulz1986}, let us first consider $H_-$.
%Let us fist consider $H_-$, following Schulz \cite{Schulz1986}.
%The velocity $v_-$ is well-defined as long as $D<\Delta+\frac{\pi}{2}$.
At the Gaussian fixed point, i.e., when the three nonlinear terms in $H_-$ are small,
the scaling dimensions of $\cos(\sqrt{8\pi}\phi_-)$ and
$\cos(\sqrt{2\pi}\theta_-)$ are $2K_-$ and $1/2K_-$, respectively.
%When $2K_->1$, i.e., when $D>\Delta-\frac{3}{2}\pi$
%in the lowest order in $D$ and $\Delta$
%This is the case in both the Haldane and large-$D$ phases \cite{Schultz1986},
In both the Haldane and large-$D$ phases, $\cos(\sqrt{2\pi}\theta_-)$ is a relevant operator ($2K_->1$),
and the field $\theta_-(x)$ is pinned at $\theta_-(x)=0$ (mod $\sqrt{2\pi}$) \cite{Schulz1986}.
In this case we can use the mean-field approximation
\begin{subequations}
\label{<cos>}
\begin{equation}
\langle\cos(\sqrt{2\pi}\theta_-)\rangle=C>0,
\label{<cos theta_->}
\end{equation}
where the brackets denote the ground-state average in the Haldane and large-$D$ phases.
Since the dual field $\phi_-$ is strongly fluctuating, we have
\begin{equation}
\langle\cos(\sqrt{8\pi}\phi_-)\rangle=0.
\label{<cos phi_->=0}
\end{equation}
\end{subequations}
When $K_-=\frac12$, the two operators, $\cos(\sqrt{8\pi}\phi_-)$ and
$\cos(\sqrt{2\pi}\theta_-)$, have the same scaling dimension ($=1$)
and compete with each other, giving rise to a quantum phase transition of Ising type ($c=\frac12$).
This Ising phase transition occurs at the boundary between the Haldane and N\'{e}el phases \cite{Schulz1986}.

We apply the mean-field approximation (\ref{<cos>}) to $H_\pm$ and incorporate it
into $H_+$.
We then obtain the effective Hamiltonian for the $\phi_+$ and $\theta_+$ fields,
\begin{align}
H_+'=&\int dx\,\biggl\{
\frac{v_+}{2}\biggl[\frac{1}{K_+}\left(\frac{d\phi_+}{dx}\right)^2
+K_+\left(\frac{d\theta_+}{dx}\right)^2\biggr]
\nonumber\\
&\qquad\quad
+[a^2(\Delta-D)+\tilde{b}^2C]\cos(\sqrt{8\pi}\phi_+)
\biggr\}.
\label{H_+'}
\end{align}
It was shown in Ref.~\cite{Schulz1986} that, in the Haldane and large-$D$ phases,
$\phi_+$ is pinned at different minima of the cosine potential
$\cos(\sqrt{8\pi}\phi_+)$.
% which is a relevant operator due to its scaling dimension $2K_+<1$.
%Since the scaling dimension of $\cos(\sqrt{8\pi}\phi_+)$ is $2K_+$,
%this operator is relevant when $K_+<1$, i.e., when
%\begin{equation}
%D+3\Delta>0
%\label{relevant phi_+}
%\end{equation}
%in the lowest order in $\Delta$ and $D$.
%In this case, we have the following two phases.
The Haldane phase is realized when the coefficient, $a^2(\Delta-D)+\tilde{b}^2C$,
of $\cos(\sqrt{8\pi}\phi_+)$ is positive and $\langle\cos(\sqrt{8\pi}\phi_+)\rangle<0$.
The large-$D$ phase is realized when the coefficient
of $\cos(\sqrt{8\pi}\phi_+)$ is negative and $\langle\cos(\sqrt{8\pi}\phi_+)\rangle>0$.
The phase transition between the Haldane phase and the large-$D$ phase
occurs when
\begin{equation}
D=\Delta+\frac{\tilde{b}^2C}{a^2}.
\label{Gaussian}
\end{equation}
Therefore, the low-energy theory at criticality is the Gaussian model \cite{Schulz1986}.
We note that we have included the correction due to the finite expectation value $C$,
which qualitatively explains the phase boundary obtained
in Ref.~\cite{ChenHS2003}.
%in contrast to Ref.~\cite{Schulz1986}.
%he did not include the correction due to the finite expectation value $C$.
%His $J_z$ is $-\Delta$.

%It should be noted that the condition (\ref{relevant phi_+}) for gapping $\phi_+$ does not agree %with the true phase boundary of the gapless XY1 and XY2 phases
%determined in Ref.~\cite{ChenHS2003}.
%{\bf (*** Need consideration ***)}
%  (Any solution?)
Incidentally, in the XY1 and XY2 phases the cosine term $\cos(\sqrt{8\pi}\phi_+)$
is irrelevant and $H_+$ is again renormalized to the Gaussian model \cite{Schulz1986}.
The phase transition between the Haldane and XY1 phases is the
Berezinskii-Kosterlitz-Thouless (BKT) transition, which is located on
the $\Delta=0$ line, where a nontrivial SU(2) symmetry exists \cite{KitazawaHN2003}.
The multicritical point located at the crossing point of the phase boundaries of
the XY1, XY2, Haldane, and N\'{e}el phases has recently been studied in
Ref.~\cite{ShiraishiN2025}.

\subsection{Correlation functions in infinite spin chains}\label{subsec:corr_thermo}

In this section, we introduce bosonic representations of various operators and discuss
their correlation functions in the spin-1 chain of infinite length
at the phase transition between the Haldane phase
and the large-$D$ phase,
where the condition (\ref{Gaussian}) is satisfied.
This quantum criticality is described by the $c=1$ Gaussian theory
forming a TLL,
%$H_+'$ (\ref{H_+'}) without the $\cos(\sqrt{8\pi}\phi_+)$ term, i.e.,
\begin{equation}
H_+^{(0)}=\frac{v_+}{2}\int dx\left[
\frac{1}{K_+}\left(\frac{d\phi_+}{dx}\right)^2+K_+\left(\frac{d\theta_+}{dx}\right)^2
\right],
\label{H_+^0}
\end{equation}
together with the gapped Hamiltonian $H_-$, where $\theta_-$ is pinned
as in Eq.~(\ref{<cos>}).
The TLL parameter $K_+$ controlling correlation exponents is a renormalized quantity
whose precise dependence on $\Delta$ and $D$ is not known.
We will obtain $K_+$ numerically by fitting correlation functions computed from
DMRG simulations.

We use Eqs.\ (\ref{S=s1+s2}) and (\ref{s^z_n,j}) to write the spin operator $S_j^z$ as
\begin{align}
S_j^z=&\,
s_{1,j}^z+s_{2,j}^z
\nonumber\\
%=&\frac{1}{\sqrt\pi}\left(\frac{d\phi_1}{dx}+\frac{d\phi_2}{dx}\right)
%\nonumber \\
%&+(-1)^ja\left(\sin(\sqrt{4\pi}\phi_1)+\sin(\sqrt{4\pi}\phi_2)\right)
%\nonumber\\
=&\,
\sqrt{\frac{2}{\pi}}\frac{d\phi_+}{dx}
+2a(-1)^j\sin(\sqrt{2\pi}\phi_+)\cos(\sqrt{2\pi}\phi_-).
\label{S_j^z bosonization}
\end{align}
The second, oscillating term yields short-range correlation
because correlation functions of the strongly fluctuating
$\cos(\sqrt{2\pi}\phi_-)$ should decay exponentially.
If there were $\sin(\sqrt{16\pi}\phi_{1,2})$ in Eq.~(\ref{s^z_n,j}),
then the resulting $\cos(\sqrt{8\pi}\phi_-)$ factor could be canceled by
the same $\cos(\sqrt{8\pi}\phi_-)$ term in $H_-$ or $H_\pm$.
However, as discussed in Ref.~\cite{HikiharaF2004},
the higher-order oscillating contributions to $s_{n,j}^z$
contain only the terms of the form $\sin[(2l+1)\sqrt{4\pi}\phi_n]$,
which do not include $\cos(\sqrt{8\pi}\phi_-)$. 
Therefore, the algebraic correlations are present only in the
uniform part in the two-point correlation function of $S_j^z$,
%\begin{equation}
%S_j^z=\sqrt{\frac{2}{\pi}}\frac{d\phi_+}{dx}
%+\ldots.
%\label{S_j^z bosonization}
%\end{equation}
%Its correlation function at the thermodynamic limit is given by
\begin{equation}
\langle S_j^z S_{j+r}^z\rangle
=-\frac{K_+}{\pi^2}\frac{1}{r^2}+(-1)^r A'_z\frac{e^{-|r|/\xi}}{|r|^{K_+}} +\ldots,
\end{equation}
where $A'_z$ is a nonuniversal constant, and $\xi$ is a correlation length that is related
to the gap in the $(\phi_-,\theta_-)$ sector.
%Once again we note that oscillating terms $\propto(-1)^r$ decay exponentially.
%The absence of oscillating power-law decaying terms is similar to the behavior
%of correlation functions in the critical phase of
%the spin-$\frac12$ two-leg ladder just below the saturation magnetic field 
%($M\to\frac12$) \cite{HikiharaF2001} (cite Sato as well).

The spin raising operator $S_j^+=S_j^x+iS_j^y$ is bosonized as
\begin{align}
S_j^+=&\, s_{1,j}^+ + s_{2,j}^+
\nonumber\\
%=&\,
%b(-1)^j\left(e^{i\sqrt{\pi}\theta_1}+e^{i\sqrt{\pi}\theta_2}\right)
%\nonumber \\
%&+\tilde{b}\left(e^{i\sqrt{\pi}\theta_1}\sin(\sqrt{4\pi}\phi_1)
% +e^{i\sqrt{\pi}\theta_2}\sin(\sqrt{4\pi}\phi_2)\right)
%\nonumber\\
=&\,
2b\widetilde{C}(-1)^j e^{i\sqrt{\pi/2}\,\theta_+}
\nonumber\\
&
+2\tilde{b}\widetilde{C}e^{i\sqrt{\pi/2}\,\theta_+}\sin(\sqrt{2\pi}\phi_+)\cos(\sqrt{2\pi}\phi_-),
\label{S_j^+ bosonization}
\end{align}
where we have
%dropped the terms containing $\phi_-$ and 
defined $\widetilde{C}=\langle\cos(\sqrt{\pi/2}\,\theta_-)\rangle$.
The bosonized form of $S_j^-=S_j^x-iS_j^y$ is obtained by taking conjugation.
Using the same line of reasoning as for the longitudinal spin $S^z_j$, we deduce that 
the second term $\propto\cos(\sqrt{2\pi}\phi_-)$ in Eq.\ (\ref{S_j^+ bosonization})
produces exponentially decaying short-range correlations,
%It is worth noting that, in constast to the $S^z_j$ operator, only the staggered terms %remains in the $S^+_j$ operator since its uniform terms contain the $\phi_-$ field.
and that the transverse spin correlation function exhibits algebraic decay only in oscillating
components.
The resultant correlation function has the form,
\begin{equation}
\langle S_j^+ S_{j+r}^-\rangle=
(-1)^{r}\frac{A_\perp}{|r|^{1/(4K_+)}}
+A'_\perp \frac{e^{-|r|/\xi}}{r^p} 
+\ldots,
\end{equation}
where $A_\perp$ and $A'_\perp$ are constants,
and $p=K_++(4K_+)^{-1}$.
%Uniform components decay exponentially, as in
%the spin-$\frac12$ two-leg ladder near saturation field ($M\to\frac12$) \cite{HikiharaF2001}.
It is interesting to note that the inverse of the exponent $1/(4K_+)$ of the leading term
of $\langle S_j^+S_{j+r}^-\rangle$ does not appear as an exponent in $\langle S_j^z S_{j+r}^z\rangle$,
but it does so in the correlation function of the squared spin operator, $\langle(S_j^z)^2(S_{j+r}^z)^2\rangle$, discussed below.

Next, we discuss the operators for the exchange interactions defined as
\begin{subequations}
\label{eq:Odim_def}
\begin{align}
\mathcal{O}_{\rm d}^{xy}(j) &= \frac{1}{4} \left( S_j^+S_{j+1}^-+S_j^-S_{j+1}^+ \right),
\label{eq:Odxy_def} \\
\mathcal{O}_{\rm d}^z(j) &= S_j^zS_{j+1}^z,
\label{eq:Odz_def}
\end{align}
\end{subequations}
and the squared longitudinal-spin operator $(S^z_j)^2$.
We call the operators in Eq.\ (\ref{eq:Odim_def}) the dimer operators in the following.
The bosonized forms of $\mathcal{O}_{\rm d}^{xy}(j)$, $\mathcal{O}_{\rm d}^z(j)$, and $(S^z_j)^2$ can be readily obtained from the result of
the previous section as
\begin{subequations}
\label{eq:boson_repre_dimer}
\begin{align}
\mathcal{O}_{\mathrm{d}}^{xy}(j) =&\,\frac{1}{4}\left[\left(\frac{d\phi_+}{dx}\right)^2
+\left(\frac{d\theta_+}{dx}\right)^2\right]
\nonumber \\
&\,
~~~+\frac{1}{2}\tilde{b}^2C\cos(\sqrt{8\pi}\phi_+)
+\mathrm{const},
\label{eq:boson_repre_Odim_xy}\\
\mathcal{O}_{\mathrm{d}}^z(j)=&\,\frac{3}{\pi}\left(\frac{d\phi_+}{dx}\right)^2
+a^2\cos(\sqrt{8\pi}\phi_+)
+\mathrm{const},
\label{eq:boson_repre_Odim_z}
\end{align}
\end{subequations}
\begin{align}
(S_j^z)^2=&\,
\frac{1}{\pi}\left(\frac{d\phi_+}{dx}\right)^2-a^2\cos(\sqrt{8\pi}\phi_+)
+\mathrm{const}.
\label{eq:boson_repre_Szj2}
\end{align}
Here, we have discarded the operators involving the $\phi_-$ field.
Their correlation functions have the same form,
\begin{eqnarray}
\mathcal{N}^\mu_{\rm d}(r) &\equiv&
\langle\mathcal{O}_{\rm d}^\mu(j) \mathcal{O}_{\rm d}^\mu(j+r)\rangle
- \langle\mathcal{O}_{\rm d}^\mu(j)\rangle
\langle\mathcal{O}_{\rm d}^\mu(j+r)\rangle
\nonumber\\
&=&\frac{A_{\rm d}^\mu}{|r|^{4K_+}}+\frac{\tilde{A}_{\rm d}^\mu}{r^4}+\ldots,
\label{dimer correlations} \\
\mathcal{N}_{\rm sq}(r) &\equiv&
\langle (S_j^z)^2 (S_{j+r}^z)^2\rangle - \langle(S_j^z)^2\rangle\langle(S_{j+r}^z)^2\rangle
\nonumber \\
&=&\frac{A_2}{|r|^{4K_+}}+\frac{\tilde{A}_2}{r^4}+\ldots,
\label{<(S_j^z)^2(S_j+r^z)^2>}
\end{eqnarray}
where $A_2$, $\tilde{A}_2$, $A_{\rm d}^\mu$, and $\tilde{A}_{\rm d}^\mu$ ($\mu=xy,z$) are nonuniversal constants.
In these equations, only the uniform components show algebraic decay, and exponentially decaying oscillating terms $\propto(-1)^r$ are ignored.

Finally, the squared spin-raising operator $(S_j^+)^2$ is bosonized as
\begin{equation}
(S_j^+)^2=2b^2e^{i\sqrt{2\pi}\theta_+}
-\tilde{b}^2e^{i\sqrt{2\pi}\theta_+}\cos(\sqrt{8\pi}\phi_+),
\end{equation}
where we have dropped the terms
that contain the $\phi_-$ field for simplicity.
Its correlation function is given by
\begin{equation}
\langle(S_j^+)^2 (S_{j+r}^-)^2\rangle
=\frac{A_\perp''}{|r|^{1/K_+}}+\frac{A_\perp'''}{|r|^{4K_++\frac{1}{K_+}}}+\ldots,
\end{equation}
where $A_\perp''$ and $A_\perp'''$ are constants,
and exponentially decaying oscillating terms are neglected.

\subsection{Correlation functions in chains under open boundary conditions}\label{subsec:corr_openchain}

In this section, we present the correlation functions in finite open chains with $L$ spins at the Gaussian criticality between the Haldane and large-$D$ phases.
The ground-state expectation values of correlation functions are obtained by taking $H_+^{(0)}$ in Eq.\ (\ref{H_+^0}) as the effective theory and imposing the open boundary conditions on the bosonic fields.
See Appendix\ \ref{app:finite_open_chain} for the derivation.
The formulas in the following will be used in Sec.\ \ref{sec:Num} to fit the numerical data obtained from DMRG calculations.

The one-point functions of the dimer operators $\mathcal{O}_{\rm d}^\mu(j)$ ($\mu = z, xy$) and the squared spin operator $(S^z_j)^2$ have the same functional form up to nonuniversal constants,
\begin{align}
\langle \mathcal{O}_{\rm d}^\mu(j)\rangle &= c_0^\mu + \frac{c_1^\mu}{[f(2j+1)]^{2K_+}} + \frac{c_2^\mu}{[f(2j+1)]^2}+\ldots,
%\nonumber \\
\label{eq:form_dimeroperator}\\
\langle (S^z_j)^2\rangle &= c'_0 + \frac{c'_1}{[f(2j)]^{2K_+}} + \frac{c'_2}{[f(2j)]^2}+\ldots.
\label{eq:form_Sz2operator}
\end{align}
%where $x$ is the center position of the operators, i.e., $x=j + \frac{1}{2}$ for $\mathcal{O}(j)= \mathcal{O}_{\rm d}^z(j), \mathcal{O}_{\rm d}^{xy}(j)$ and $x=j$ for $\mathcal{O}(j)= (S^z_j)^2$.
Here, we have defined the function $f(x)$ as
\begin{eqnarray}
f(x) = \frac{2(L+1)}{\pi} \sin\left( \frac{\pi |x|}{2(L+1)} \right).
\label{eq:fx}
\end{eqnarray}
The argument of $f(x)$ in Eq.\ (\ref{eq:form_dimeroperator}) reflects the fact that
$\mathcal{O}_\mathrm{d}^\mu(j)$ is defined on the link between the sites $j$ and $j+1$.
We note that in addition to the terms appearing in Eqs.\ (\ref{eq:form_dimeroperator}) and (\ref{eq:form_Sz2operator}), there are staggered components decaying exponentially from the open boundaries.

The two-point correlation functions of the longitudinal-spin, transverse-spin, dimer, and squared-spin operators are obtained as
\begin{eqnarray}
&&\langle S^z_j S^z_k \rangle =
-\frac{K_+}{\pi^2} \left\{
\frac{1}{[f(j-k)]^2} + \frac{1}{[f(j+k)]^2} \right\},
\label{eq:form_SzSzcor} \\
&&\langle S^+_j S^-_k \rangle = 
(-1)^{j-k} A_\perp \frac{[f(2j)f(2k)]^{\frac{1}{8K_+}}}{[f(j-k)f(j+k)]^{\frac{1}{4K_+}}},
\label{eq:form_S+S-cor} \\
&&\mathcal{N}^{\mu}_{\rm d}(j,k) 
\nonumber \\
&&~~\equiv
\langle \mathcal{O}_{\rm d}^\mu(j) \mathcal{O}_{\rm d}^\mu(k) \rangle -  \langle \mathcal{O}_{\rm d}^\mu(j) \rangle \langle \mathcal{O}_{\rm d}^\mu(k) \rangle
\nonumber \\
&&~~= \frac{A_{\rm d}^\mu}{[f(2j+1)f(2k+1)]^{2K_+}}
\nonumber \\
&&~~~~\times
\!\left\{ \!
\left[\frac{f(j+k+1)}{f(j-k)}\right]^{4K_+} \! + \left[\frac{f(j-k)}{f(j+k+1)}\right]^{4K_+} \!
-2
\right\} \! ,
\nonumber \\
\label{eq:form_dimercorr} \\
&&\mathcal{N}_{\rm sq}(j,k) 
\nonumber \\
&&~~\equiv \langle (S^z_j)^2 (S^z_k)^2 \rangle -  \langle (S^z_j)^2 \rangle \langle (S^z_k)^2 \rangle
\nonumber \\
&&~~= \frac{A_2}{[f(2j)f(2k)]^{2K_+}}
\nonumber \\
&&~~~~~\times 
\!\left\{ \!
\left[\frac{f(j+k)}{f(j-k)}\right]^{4K_+} \! + \left[\frac{f(j-k)}{f(j+k)}\right]^{4K_+} \!
-2 
\right\} \! .
\label{eq:form_Sz2corr}
\end{eqnarray}
In Eqs.\ (\ref{eq:form_dimercorr}) and (\ref{eq:form_Sz2corr}), we have omitted the contributions from the terms proportional to $\left( \frac{d\phi_+}{dx}\right)^2$ and $\left( \frac{d\theta_+}{dx}\right)^2$ in Eqs.\ (\ref{eq:boson_repre_dimer}) and (\ref{eq:boson_repre_Szj2}), which lead to the term of $\sim 1/r^4$ in the correlation function in the limit $L \to \infty$, since their contributions are only subleading.
(See also the result of $K_+$ in Sec.\ \ref{subsec:exponent}.)

Before closing this section, we discuss the boundary correlation functions in semi-infinite chains.
In the limit $L\to\infty$ and for finite $x$,
Eqs.~(\ref{eq:form_dimeroperator})
and (\ref{eq:form_Sz2operator}) become
\begin{align}
\langle\mathcal{O}_\mathrm{d}^\mu(j)\rangle=&\,
c_0^\mu+\frac{c_1^\mu}{|2j+1|^{2K_+}}+\frac{c_2^\mu}{|2j+1|^2}+\ldots,
\\
\langle(S_j^z)^2\rangle=&\,
c_0'+\frac{c_1'}{|2j|^{2K_+}}+\frac{c_2'}{|2j|^2}+\ldots.
\end{align}
Here, contributions that are exponentially localized near the boundary $j=1$ are neglected.
The exponent $2K_+$ is a half of the exponent $4K_+$ of the correlation functions
in the bulk in Eqs.\ (\ref{dimer correlations}) and (\ref{<(S_j^z)^2(S_j+r^z)^2>}).

\subsection{Effect of bond alternation}\label{subsec:bondalternation}

We have seen in the previous sections that the power-law decaying correlations 
in $\langle S_j^z S_k^z\rangle$ and $\langle S_j^+ S_k^-\rangle$
exist only in the uniform and alternating terms, respectively.
This is a characteristic feature of the model (\ref{eq:Ham}) at
the critical points between the Haldane and large-$D$ phases.
In this section we argue that both alternating and uniform components acquire
power-law correlations when we perturb the Hamiltonian $H$ in Eq.~(\ref{eq:Ham})
by adding weak bond alternation,
\begin{equation}
H_\delta=\delta \sum_j (-1)^{j-1} (S_j^xS_{j+1}^x + S_j^yS_{j+1}^y +\Delta S_j^zS_{j+1}^z).
\label{H_delta}
\end{equation}
From Eq.\ (\ref{h_n(j)}) we find the bosonization of $H_\delta$ yields
\begin{equation}
\widetilde{H}_\delta=c_\delta \delta\int dx \cos(\sqrt{2\pi}\phi_+)\cos(\sqrt{2\pi}\phi_-),
\label{H_delta bosonization}
\end{equation}
where $c_\delta$ is a constant.

We have discussed in Sec.~\ref{subsec:bosonization} that $H_-$ is gapped due to
the pinning of $\theta_-$ and that $\phi_-$ is strongly fluctuating.
Therefore, $\widetilde{H}_\delta$ is an irrelevant perturbation and does not
affect the critical theory qualitatively.
One obvious consequence of the addition of weak bond alternation is that
the phase boundary between the Haldane and large-$D$ phases can be slightly shifted: 
since second-order perturbation can yield the operator
$\cos(\sqrt{8\pi}\phi_+)$, the position of the critical points
will be shifted on the order $\delta^2$ for small $\delta$.

Another important consequence of the addition of weak bond alternation exists
in the correlation functions.
We note that both $S_j^z$ and $S_j^+$ operators have an operator
$\sin(\sqrt{2\pi}\phi_+)\cos(\sqrt{2\pi}\phi_-)$ in their bosonized formulas
in Eqs.\ (\ref{S_j^z bosonization}) and (\ref{S_j^+ bosonization}).
In first-order perturbation in $\widetilde{H}_\delta$,
the product of this operator and the bond alternation
operator $\cos(\sqrt{2\pi}\phi_+)\cos(\sqrt{2\pi}\phi_-)$
in Eq.\ (\ref{H_delta bosonization}) generates $\sin(\sqrt{8\pi}\phi_+)$,
which does not involve $\phi_-$ anymore.
Therefore, the bosonized form of the spin operators is effectively modified to
\begin{subequations}
\begin{align}
S_j^z=&\,
\sqrt{\frac{2}{\pi}}\frac{d\phi_+}{dx}
-\tilde{a}(-1)^j\sin(\sqrt{8\pi}\phi_+)+\ldots,
\label{eq:Sz_bosonic_BA}
\\
S_j^+=&\,
2b\widetilde{C}(-1)^j e^{i\sqrt{\pi/2}\theta_+}
+\tilde{b}' e^{i\sqrt{\pi/2}\theta_+}\sin(\sqrt{8\pi}\phi_+)+\ldots,
\end{align}
\end{subequations}
where $\tilde{a}$ and $\tilde{b}'$ are constants proportional to $\delta$.
Their bulk two-point correlation functions are also changed to
\begin{subequations}
\label{<SS> with bond alternation}
\begin{align}
\langle S_j^z S_{j+r}^z\rangle=&\,
-\frac{K_+}{\pi^2 r^2}+(-1)^r\frac{\widetilde{A}_z}{|r|^{4K_+}}+\ldots,
\label{SzSz_thermodynamic_BA} \\
\langle S_j^+ S_{j+r}^-\rangle=&\,
(-1)^r\frac{A_\perp}{|r|^{1/(4K_+)}}+\frac{\widetilde{A}_\perp}{|r|^{4K_++\frac{1}{4K_+}}}
+\ldots,
\end{align}
\end{subequations}
where $\widetilde{A}_z$ and $\widetilde{A}_\perp$ are constants
that are proportional to $\delta^2$ for small $|\delta|$.
We note that the $r$ dependence in Eqs.~(\ref{<SS> with bond alternation})
is similar to the case of the spin-$\frac12$ XXZ chain,
as observed earlier in Ref.\ \cite{EjimaYELOF2018}.

Using Eqs.\ (\ref{eq:Sz_bosonic_BA}) and (\ref{eq:mode_expansion_phi+}),
we obtain the analytic form of the longitudinal two-spin correlation function $\langle S^z_j S^z_k \rangle$
in the finite open chain as
\begin{eqnarray}
&&\langle S^z_j S^z_k \rangle
\nonumber \\
&&= -\frac{K_+}{\pi^2} \left\{ \frac{1}{[f(j-k)]^2} + \frac{1}{[f(j+k)]^2} \right\}
\nonumber \\
&&~ + \frac{\tilde{a}^2 (-1)^{j-k}}{2[f(2j)f(2k)]^{2K_+}}
\left\{ \left[ \frac{f(j+k)}{f(j-k)} \right]^{4K_+} \!
- \left[ \frac{f(j-k)}{f(j+k)} \right]^{4K_+} \right\}
\nonumber \\
&&~ - \frac{2K_+}{\pi}\tilde{a} \left\{
\frac{(-1)^j}{[f(2j)]^{2K_+}} [g(j+k)+g(j-k)]\right.
\nonumber \\
&&\left.\qquad\qquad + \frac{(-1)^k}{[f(2k)]^{2K_+}} [g(j+k)-g(j-k)] \right\},
\label{eq:SzSz_open_form_BA}
\end{eqnarray}
where 
\begin{eqnarray}
g(x) &=& \frac{\pi}{2(L+1)} \cot\left(\frac{\pi x}{2(L+1)}\right).
\label{eq:gx}
\end{eqnarray}
and $f(x)$ is defined in Eq.\ (\ref{eq:fx}).
We will use this form in Sec.\ \ref{subsec:Num_bondalternation}.

\subsection{Spin-$\frac12$ at the domain wall between the Haldane and large-$D$ phases}

The Haldane and large-$D$ phases are topologically distinct phases,
and a free spin $S=\frac12$ should exist at a domain wall between
the two phases.
Let us briefly discuss how such a spin localized at a domain wall is understood
in the effective low-energy theory;
see also Refs.~\cite{Fuji2016,LecheminantO2002} for related discussions.

Suppose that we have a domain wall at $x=0$, which separates the large-$D$ phase region for $x<0$
and the Haldane phase region for $x>0$.
This means that the coupling constant, $g=a^2(\Delta-D)+\tilde{b}^2C$, in
the effective theory $H_+'$ in Eq.\ (\ref{H_+'}) changes its sign
from a negative to a positive value as $x$ increases across the domain wall.
When the phase field $\phi_+$ is pinned at $\langle\phi_+\rangle=0$
in the large-$D$ domain ($x<0$)
and at $\langle\phi_+\rangle=\pm\sqrt{\pi/8}$ in the Haldane domain ($x>0$),
the spin localized at the domain wall
can be estimated from Eq.\ (\ref{S_j^z bosonization}) as
\begin{equation}
\Delta S^z = \sqrt{\frac{2}{\pi}}\Delta\phi_+
=\sqrt{\frac{2}{\pi}}\left(\pm\sqrt{\frac{\pi}{8}}\right)=\pm\frac{1}{2}.
\end{equation}
Therefore, a spin with either $S^z=\frac12$ or $S^z=-\frac12$ is localized at the domain wall.

Here we recall that the ground state of the Haldane phase is unique in the bulk,
since the two pinning configurations of the field at $\phi_+=\sqrt{\pi/8}$
and $\phi_+=-\sqrt{\pi/8}$ are equivalent.
This can be understood by noting that the original fields $\phi_{1,2}$ are compactified
with the radius $R=1/\sqrt{4\pi}$ as $\phi_n\equiv\phi_n+\sqrt{\pi}$,
thereby $\phi_+=(\phi_1+\phi_2)/\sqrt{2}\equiv\phi_++\sqrt{\pi/2}$.
%(Note also that the $\phi_-$ field is strongly fluctuating because of the pinning of the $\theta_-$ field.)
Similarly, the ground state of the large-$D$ phase is unique in the bulk,
and the pinning at $\phi_+=0\equiv \pm\sqrt{\pi/2}$.

%\subsection{Dynamical structure factor}\label{subsec:DSF}

\section{Numerical results}\label{sec:Num}
\subsection{Method}\label{subsec:method}

To demonstrate the validity of the effective theory derived in the previous section,
we performed numerical calculations at the critical points between the Haldane and large-$D$ phases of the model (\ref{eq:Ham}).
The critical value of the single-ion anisotropy $D_{\rm c}$ at the critical points were determined as a function of $\Delta$
using the twisted-boundary method \cite{KitazawaN1997a,KitazawaN1997b,ChenHS2000,ChenHS2008,ChenHS2003}.
Details of the analysis are presented in Appendix\ \ref{app:crt_point}.
At the critical points $(\Delta, D_{\rm c}(\Delta))$, we calculated the one-point functions and two-point correlation functions
of the spin, dimer, and squared-spin $(S^z_j)^2$ operators using the DMRG method.
In the calculations, the open boundary conditions were imposed and the system size treated was up to $L=256$.
The maximum bond dimension was $\chi=505$ and the truncation error, the average of the sum of the reduced-density-matrix weights of discarded states over the last sweep, was $1 \times 10^{-10}$ at most.
%The DMRG data obtained are accurate enough for the following arguments. {\bf (**** check needed ****)}

We perform the fitting of the correlation functions obtained numerically for $\Delta\in[0.50, 2.50]$ 
to the functional forms, Eqs.\ (\ref{eq:form_Sz2operator})-(\ref{eq:form_Sz2corr}) and Eq.\ (\ref{eq:SzSz_open_form_BA}), discussed in Secs.~\ref{subsec:corr_openchain} and \ref{subsec:bondalternation}.
In the ground-state phase diagram, the region of the Haldane phase appearing between the large-$D$ and N\'{e}el phases becomes narrower as $\Delta$ increases, and terminates at the multicritical point at $(\Delta, D) \approx (3.2, 2.9)$ \cite{ChenHS2003}.
For the transition points $(\Delta, D_{\rm c}(\Delta))$ close to the multicritical point, the system size required to realize the asymptotic behavior becomes large, making it difficult to achieve the fitting of good quality.
We have found that the fitting works very well for $\Delta \lesssim 1.50$, indicating that the low-energy effective theory correctly describes the criticality at the transition between the Haldane and large-$D$ phases, while the accuracy of the fitting becomes poor when $\Delta$ approaches the value at the multicritical point.
In the following, we mainly present the results of the fitting for the data at $\Delta = 1.50$ as a representative case.
%{\bf (**** Should we present the results for different $\Delta$? ****)}
%The results for the different $\Delta$'s are presented in Appendix\ \ref{app:Delta-dep}.

\subsection{One-point functions}\label{subsec:1p}

We first discuss the one-point functions of the dimer operators $\mathcal{O}_{\rm d}^z(j), \mathcal{O}_{\rm d}^{xy}(j)$ and the squared spin operators $(S^z_j)^2$.
We have performed the least-square fitting of the numerical data of $\langle \mathcal{O}_{\rm d}^\mu(j)\rangle$ ($\mu=z,xy$) and
$\langle (S^z_j)^2 \rangle$ to Eqs.\ (\ref{eq:form_dimeroperator}) and (\ref{eq:form_Sz2operator}),
taking $\{K_+, c^\mu_0, c^\mu_1, c^\mu_2\}$ and $\{K_+, c'_0, c'_1, c'_2\}$ as fitting parameters, respectively.
In the fitting, the data points near the open boundaries of the system must be excluded in order to avoid the influence of the exponentially-decaying staggered component, which is not included in Eqs.\ (\ref{eq:form_dimeroperator}) and (\ref{eq:form_Sz2operator}).
In practice, we have used the data of $\mathcal{O}_{\rm d}^\mu(j)$ and $(S^z_j)^2$ in the range of $l_0 + 1 \le j \le L-l_0-1$
and $l_0 +1 \le j \le L-l_0$, respectively, for three cases of $l_0 = \frac{L}{8}, \frac{5L}{32}, \frac{3L}{16}$.
We then adopt the average of the fitting parameters obtained from the three cases as the estimates of the parameters.

\begin{figure*}[t]
\begin{center}
\includegraphics[width = 50 mm]{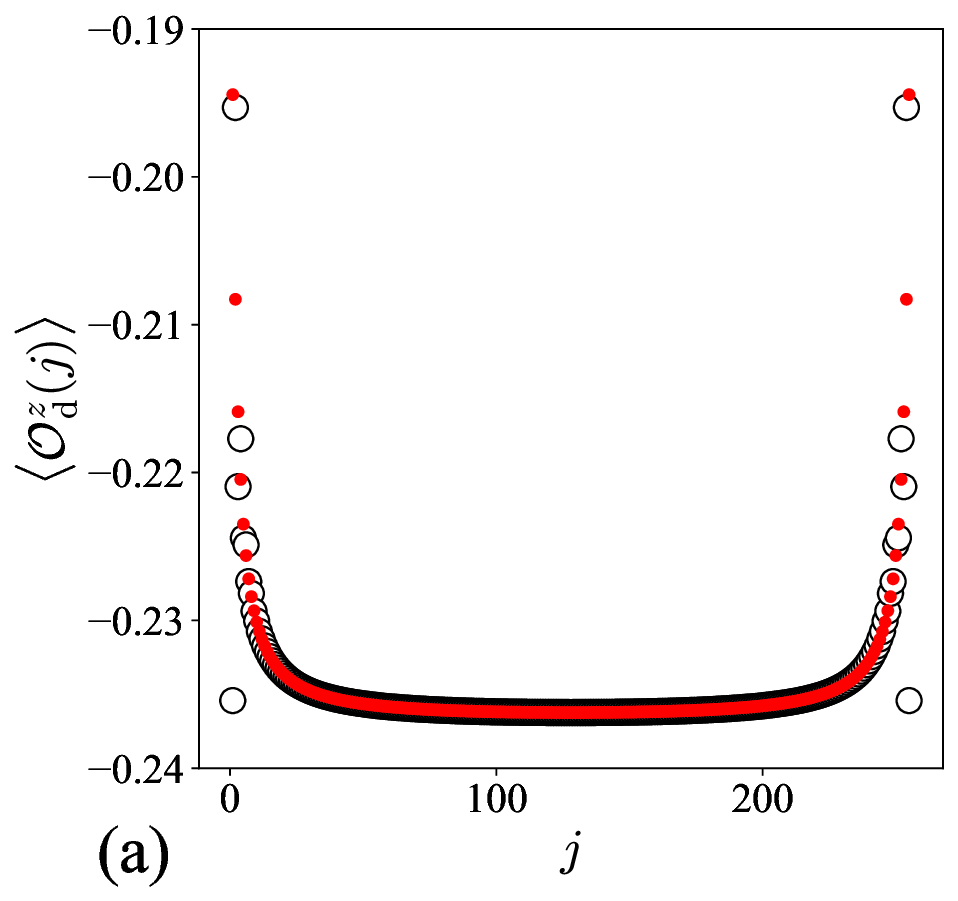}
\includegraphics[width = 50 mm]{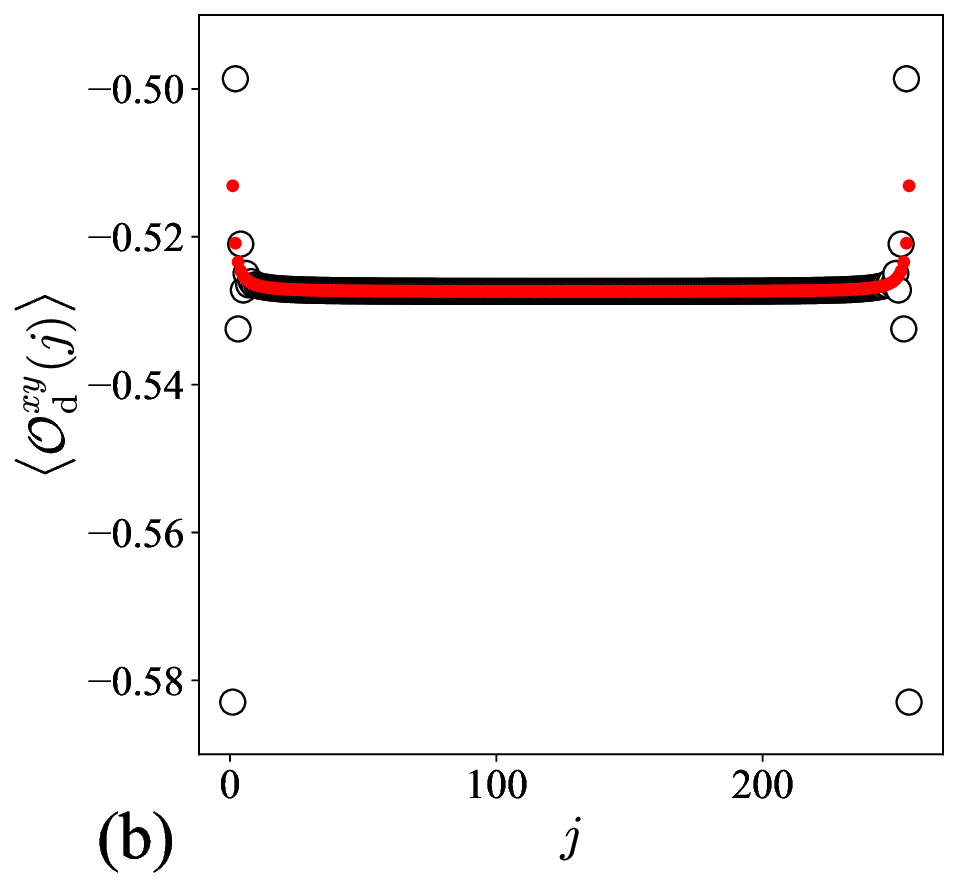}
\includegraphics[width = 50 mm]{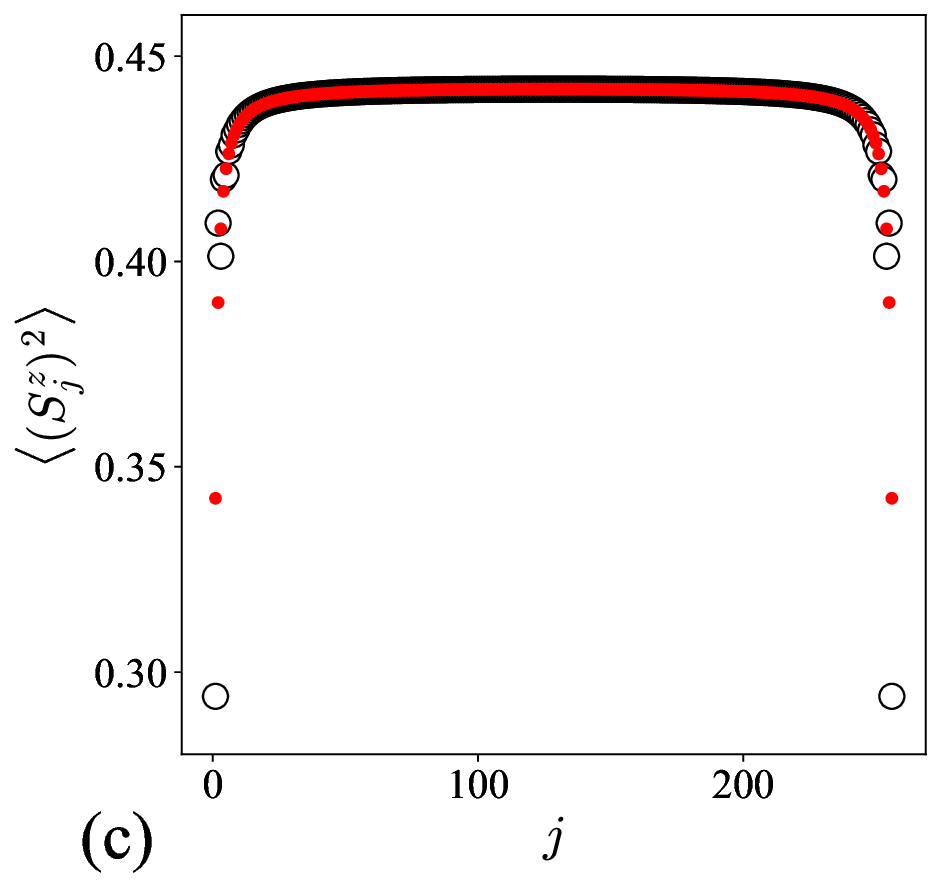}
\caption{
One-point function of (a) $\langle \mathcal{O}_{\rm d}^z(j)\rangle$, (b) $\langle \mathcal{O}_{\rm d}^{xy}(j)\rangle$, and (c) $\langle (S^z_j)^2 \rangle$ for $\Delta=1.50$ and $L=256$.
Open and solid circles respectively represent the DMRG data and their fits to the analytic forms Eqs.\ (\ref{eq:form_dimeroperator}) and (\ref{eq:form_Sz2operator}).
}
\label{fig:1p_data}
\end{center}
\end{figure*}

Figure\ \ref{fig:1p_data} presents the numerical data of $\langle \mathcal{O}_{\rm d}^z(j)\rangle$, $\langle \mathcal{O}_{\rm d}^{xy}(j)\rangle$, and $\langle (S^z_j)^2\rangle$ and the results of the fitting for $\Delta=1.50$ and $L=256$.
We find that the leading $j$ dependence comes from uniform components that decay algebraically from the open boundaries.
The staggered components decay very rapidly from the boundaries, despite having relatively large magnitudes in $\langle \mathcal{O}_{\rm d}^{xy}(j)\rangle$.
The fitting results (red solid circles) agree well with the numerical data, except for the vicinity of the boundaries.

\begin{figure*}
\begin{center}
\includegraphics[width = 50 mm]{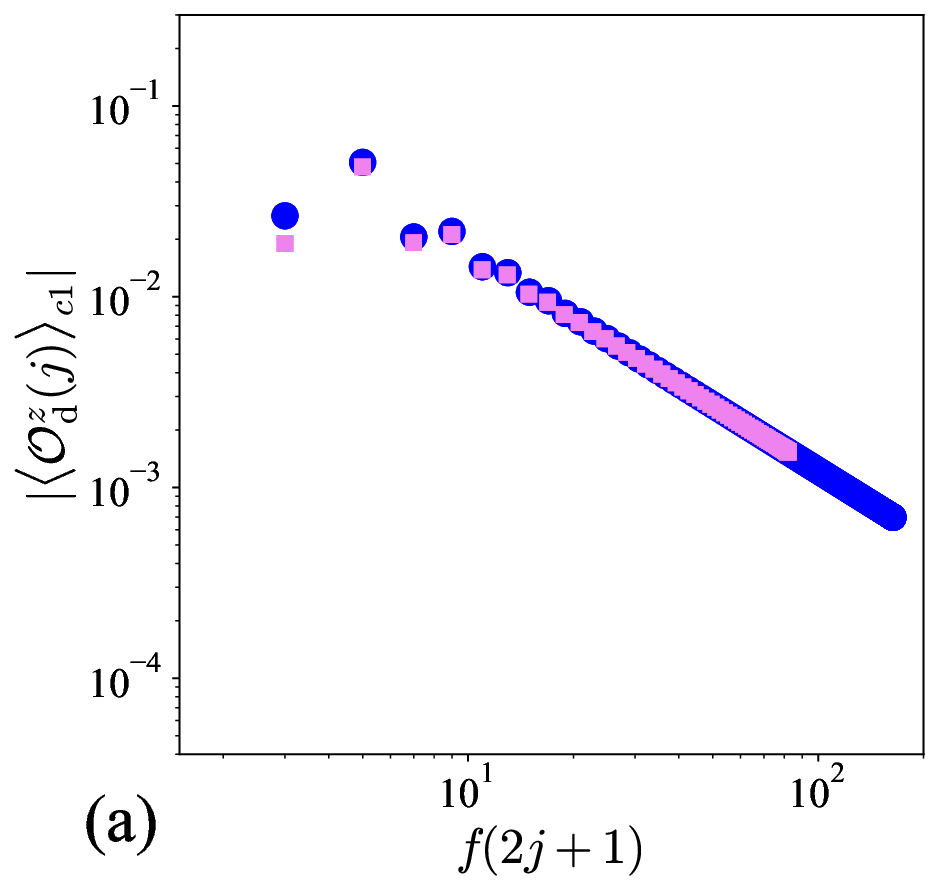}
\includegraphics[width = 50 mm]{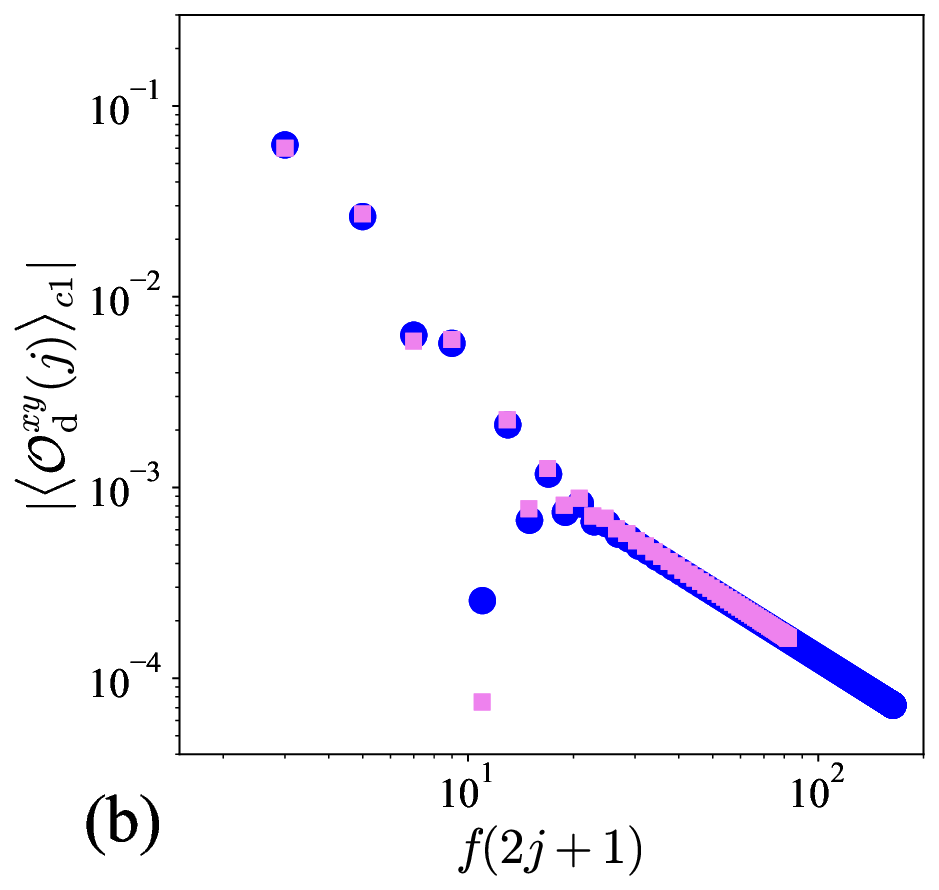}
\includegraphics[width = 50 mm]{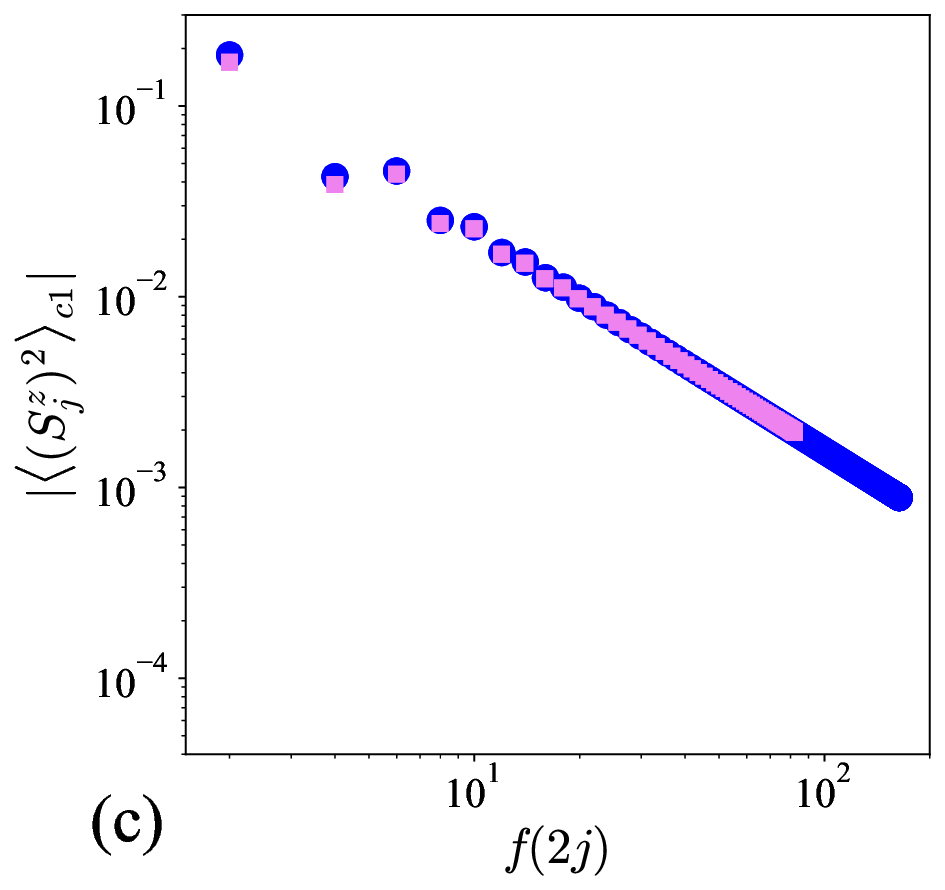}
\caption{
Leading uniform component of the one-point functions: (a) $\langle \mathcal{O}_{\rm d}^z(j)\rangle_{c1}$, (b) $\langle \mathcal{O}_{\rm d}^{xy}(j)\rangle_{c1}$, and (c) $\langle (S^z_j)^2 \rangle_{c1}$ for $\Delta=1.50$.
Circles (blue) and squares (purple) represent the data for $L=256$ and $128$, respectively.
}
\label{fig:1p-c0-c2}
\end{center}
\end{figure*}

Next, in order to extract the most slowly decaying contribution, we consider the one-point function $\langle \mathcal{O}(j)\rangle_{c1}$
for $\mathcal{O}(j)= \mathcal{O}_{\rm d}^\mu(j)$ ($\mu=z,xy$), $(S^z_j)^2$ defined by
\begin{eqnarray}
\langle \mathcal{O}_{\rm d}^\mu(j)\rangle_{c1} &=& \langle \mathcal{O}_{\rm d}^\mu(j)\rangle - c^\mu_0 - \frac{c^\mu_2}{[f(2j+1)]^2},
\label{eq:dimeroperator_c1} \\
\langle (S^z_j)^2 \rangle_{c1} &=& \langle (S^z_j)^2 \rangle - c'_0 - \frac{c'_2}{[f(2j)]^2},
\label{eq:Sz2operator_c1}
\end{eqnarray}
where $c^\mu_0$, $c^\mu_2$, $c'_0$, and $c'_2$ are the estimates obtained from the fitting of $\langle \mathcal{O}(j)\rangle$.
Note that $\langle \mathcal{O}(j)\rangle_{c1}$ corresponds to the term with the coefficient $c^\mu_1$ or $c'_1$ decaying with the exponent $2K_+$ in Eqs.\ (\ref{eq:form_dimeroperator}) and (\ref{eq:form_Sz2operator}).
Figure\ \ref{fig:1p-c0-c2} show the data of $\langle \mathcal{O}_{\rm d}^\mu(j)\rangle_{c1}$ and $\langle (S^z_j)^2\rangle_{c1}$ as functions of $f(2j+1)$ and $f(2j)$, respectively.
We can see in the figure that the plots of $\langle \mathcal{O}(j)\rangle_{c1}$ for $L=128$ and $256$ lie on the same straight line in a log-log scale, indicating that the fitting works very well.
The estimates of the decay exponent $2K_+$ from the fitting of the data for $L=256$ are $2K_+ = 1.143, 1.158$, and $1.144$ for $\mathcal{O}_{\rm d}^z(j), \mathcal{O}_{\rm d}^{xy}(j)$, and $(S^z_j)^2$, respectively, which are in good agreement with each other.
The $\Delta$ dependence of the exponent $K_+$ will be discussed later in this section.

\begin{figure*}
\begin{center}
\includegraphics[width = 50 mm]{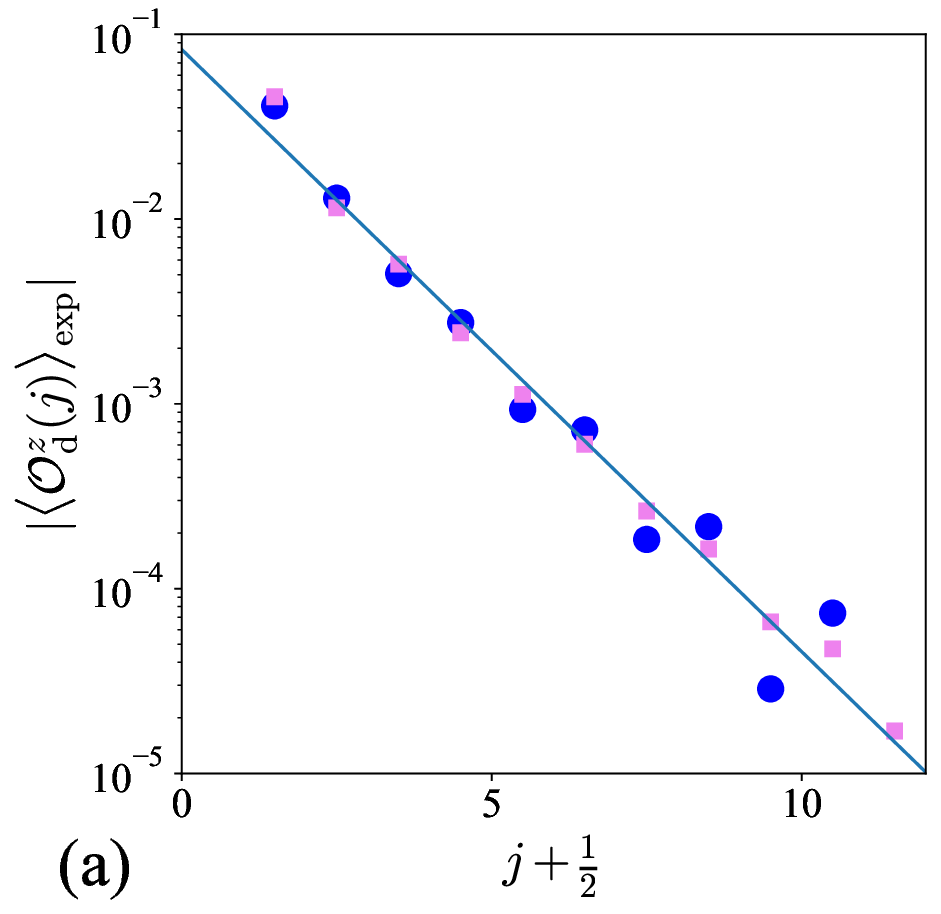}
\includegraphics[width = 50 mm]{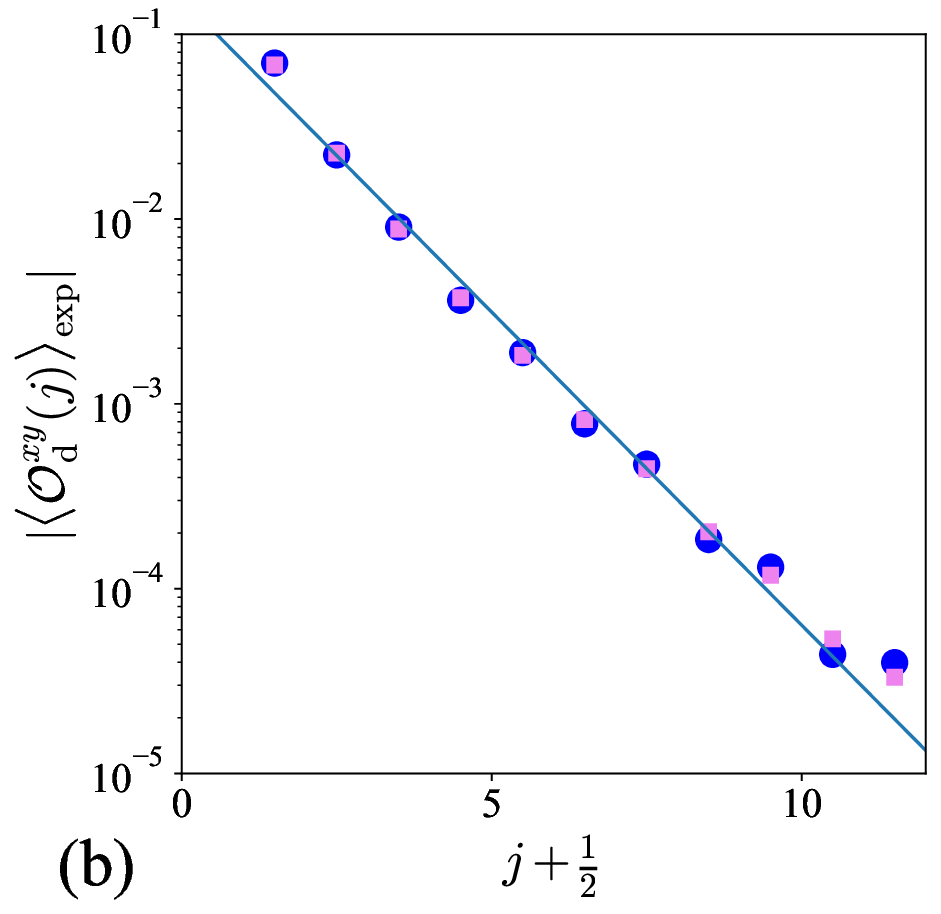}
\includegraphics[width = 50 mm]{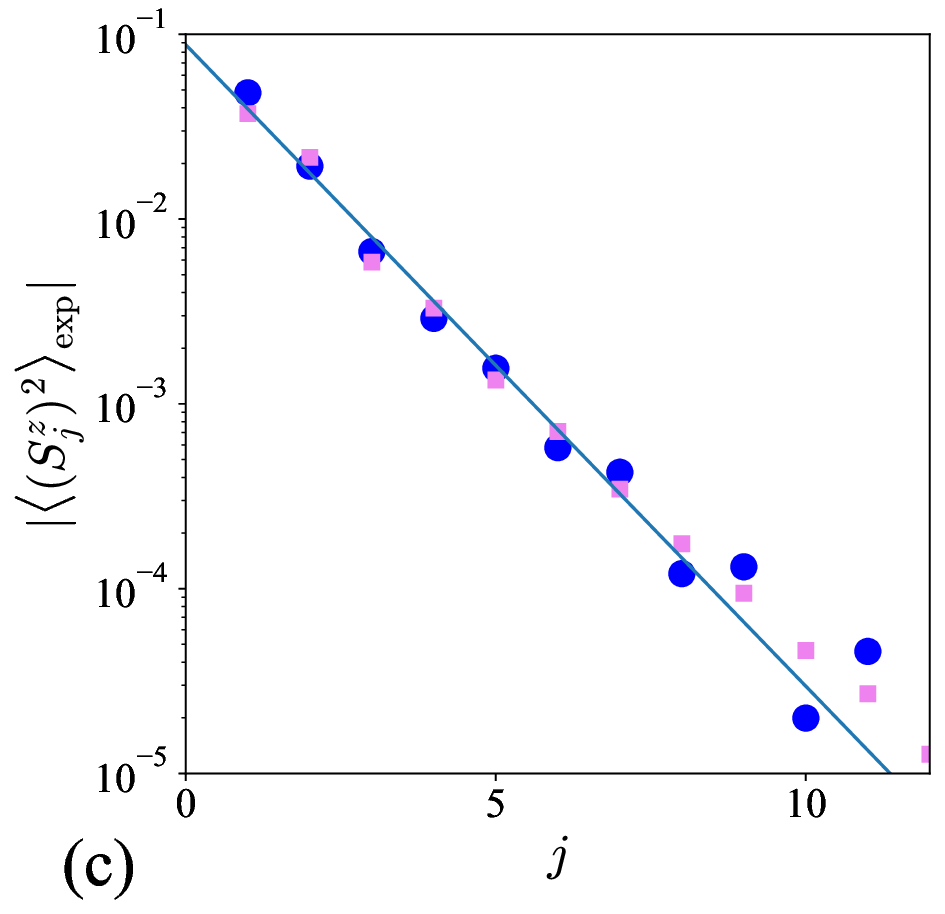}
\caption{
Exponentially-decaying component of one-point functions; (a) $\langle \mathcal{O}_{\rm d}^z(j)\rangle_{\rm exp}$, (b) $\langle \mathcal{O}_{\rm d}^{xy}(j)\rangle_{\rm exp}$, and (c) $\langle (S^z_j)^2\rangle_{\rm exp}$, for $\Delta=1.50$.
The absolute values of the data near the open boundary, $j \lesssim 12$, are plotted as a function of the center position of the operators, {\it i.e.}, $j+\frac{1}{2}$ for the dimer operators and $j$ for the squared-spin operator.
Circles (blue) and squares (purple) represent the data for $L=256$ and $128$, respectively.
The solid line shows the result of fitting to the exponentially-decaying form.
}
\label{fig:1p-c0-c1-c2}
\end{center}
\end{figure*}

Finally, we discuss the exponentially-decaying components in the one-point functions.
To examine its bahavior, we consider $\langle \mathcal{O}(j)\rangle_{\rm exp}$ defined by
\begin{eqnarray}
&&\langle \mathcal{O}_{\rm d}^\mu(j)\rangle_{\rm exp} 
\nonumber \\
&&~~= \langle \mathcal{O}_{\rm d}^\mu(j)\rangle - c^\mu_0 - \frac{c^\mu_1}{[f(2j+1)]^{2K_+}} - \frac{c^\mu_2}{[f(2j+1)]^2},
\nonumber \\
\label{eq:dimeroperator_exp} \\
&&\langle (S^z_j)^2\rangle_{\rm exp} 
\nonumber \\
&&~~= \langle (S^z_j)^2\rangle - c'_0 - \frac{c'_1}{[f(2j)]^{2K_+}} - \frac{c'_2}{[f(2j)]^2}.
\nonumber \\
\label{eq:Sz2operator_exp}
\end{eqnarray}
For the parameters $\{c^\mu_0, c^\mu_1, c^\mu_2\}$ or $\{c'_0, c'_1, c'_2\}$ and $K_+$, we used the values estimated from the fitting.
% of $\langle \mathcal{O}_{\rm d}^z(j)\rangle$, $\langle \mathcal{O}_{\rm d}^{xy}(j)\rangle$, and $\langle (S^z_j)^2\rangle$.
This quantity $\langle \mathcal{O}(j)\rangle_{\rm exp}$ should correspond to the staggered components decaying exponentially from the open boundaries, which are omitted in the analytic forms of Eqs.\ (\ref{eq:form_dimeroperator}) and (\ref{eq:form_Sz2operator}).
In Fig.\ \ref{fig:1p-c0-c1-c2}, we plot the data of $\langle \mathcal{O}(j)\rangle_{\rm exp}$ for $\mathcal{O}(j) = \mathcal{O}_{\rm d}^z(j), \mathcal{O}_{\rm d}^{xy}(j), (S^z_j)^2$ at $\Delta=1.50$ in the vicinity of the open boundary at $j=1$.
We find that the staggered boundary components decay exponentially, as expected from the effective theory.
We have also estimated the correlation length $\xi$ from the slope of the solid lines in the semi-log plots, which will be discussed later.

\subsection{Two-point Correlation functions}\label{subsec:cor}

In this section, we discuss the two-point correlation functions.
The correlation functions were calculated for the two points $(j,k)$ located at almost equal distances from the center position of the finite chain, i.e., with $j+k=L$ or $L+1$.
The fitting forms for the correlation functions derived in Sec.\ \ref{subsec:corr_openchain} do not include the effect of higher-order terms and the exponentially-decaying components, which can be sizable at short distances and near the open boundaries.
To avoid the influences of those contributions, we perform the fitting using only the data within the range $r_{\rm s} \le |j-k| \le L-r_{\rm b}$ for the three sets of $(r_{\rm s}, r_{\rm b})= (\frac{L}{32}, \frac{L}{8}), (\frac{3L}{64}, \frac{5L}{32}), (\frac{L}{16}, \frac{3L}{16})$.
We then take the average of the fitting parameters obtained from the three fittings to estimate their values.

\begin{figure}
\begin{center}
\includegraphics[width = 60 mm]{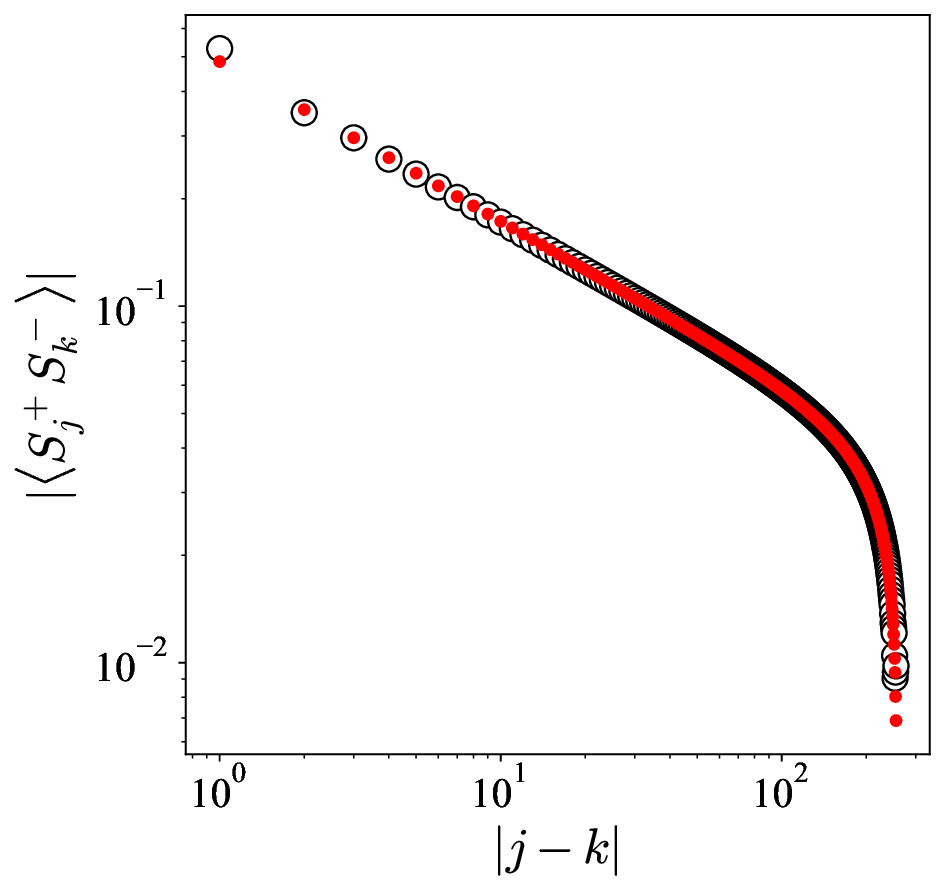}
\caption{
Transverse two-spin correlation function $\langle S^+_j S^-_k \rangle$ for $\Delta=1.50$ and $L=256$.
The absolute values are plotted as a function of $|j-k|$.
Open and solid circles respectively represent the DMRG data and their fits to the analytic form Eq.\ (\ref{eq:form_S+S-cor}).
}
\label{fig:S+S-_fits}
\end{center}
\end{figure}

Figure\ \ref{fig:S+S-_fits} presents the numerical data of the transverse two-spin correlation functions $\langle S^+_j S^-_k \rangle$ and their fits to Eq.\ (\ref{eq:form_S+S-cor}) for $\Delta=1.50$ and $L=256$.
The fitting was performed by taking the exponent $K_+$ and the amplitude $A_\perp$ as the fitting parameters.
The fitting results are in excellent agreement with the numerical data, demonstrating the validity of the analytic form.

\begin{figure}
\begin{center}
\includegraphics[width = 60 mm]{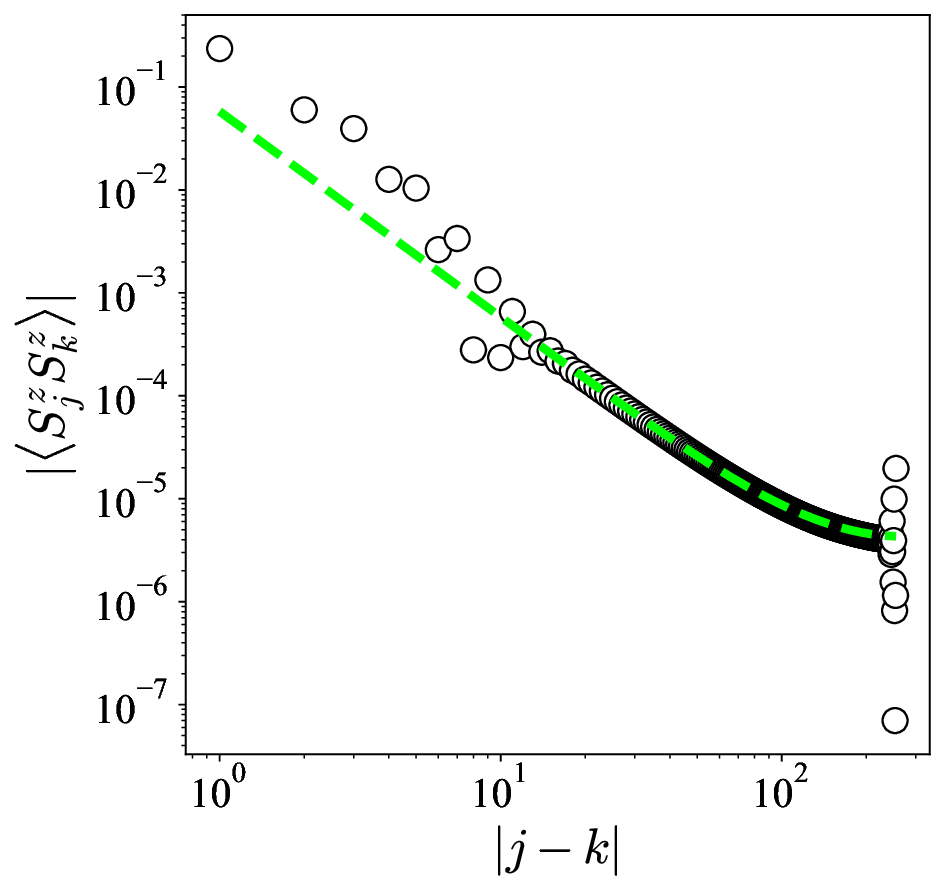}
\caption{
Longitudinal two-spin correlation function $\langle S^z_j S^z_k \rangle$ for $\Delta=1.50$ and $L=256$.
The absolute values are plotted as a function of $|j-k|$.
The open circles represent the DMRG data.
The dashed curve represents the analytic form Eq.\ (\ref{eq:form_SzSzcor}) with $K_+$ obtained from the fitting of $\langle \mathcal{O}^z_{\rm d}(j) \rangle$.
}
\label{fig:SzSz_fits}
\end{center}
\end{figure}

Next, Fig.\ \ref{fig:SzSz_fits} shows the numerical data of the longitudinal two-spin correlation function $\langle S^z_j S^z_k \rangle$ for $\Delta=1.50$ and $L=256$.
%together with the analytic form Eq.\ (\ref{eq:form_SzSzcor}).
Here, we note that $K_+$ is the only parameter in the analytic form of Eq.\ (\ref{eq:form_SzSzcor}) for this correlation function.
In Fig.\ \ref{fig:SzSz_fits}, we plot Eq.\ (\ref{eq:form_SzSzcor}) using the value of $K_+$ obtained from the fitting of $\langle \mathcal{O}^z_{\rm d}(j) \rangle$ in the prevoius section.
The analytic form agrees with the numerical data very well except for the exponentially-decaying staggered components at short distances and near the open boundaries, supporting the validity of the effective theory.

\begin{figure*}
\begin{center}
\includegraphics[width = 50 mm]{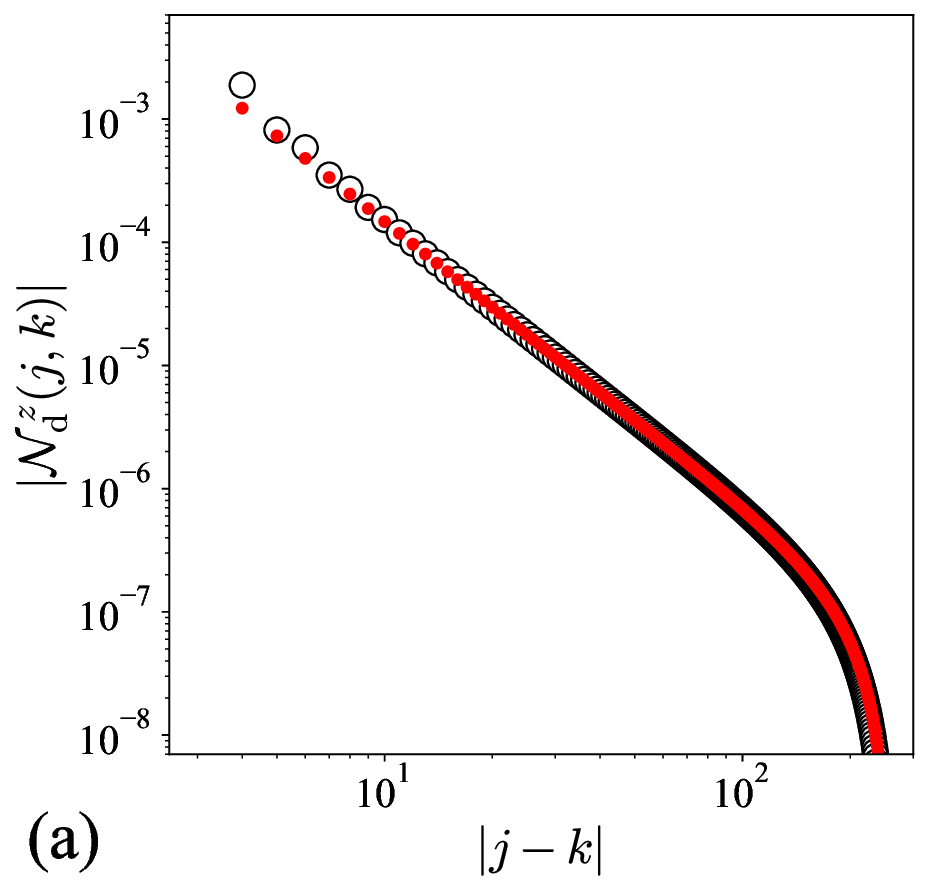}
\includegraphics[width = 50 mm]{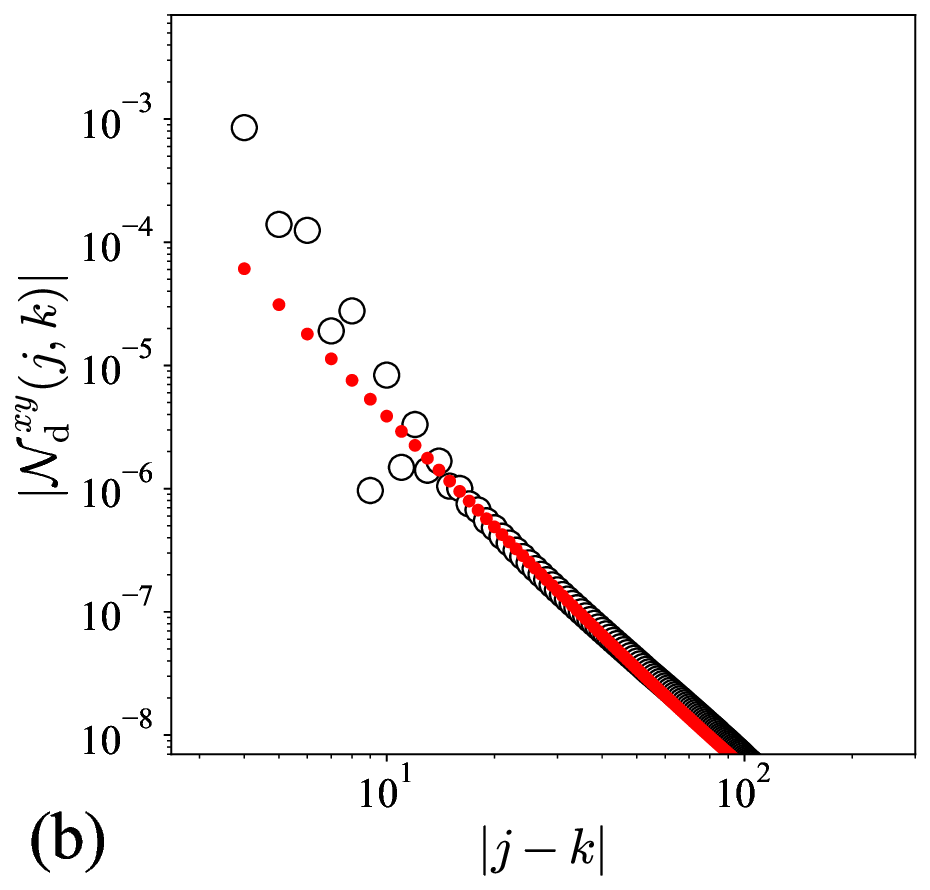}
\includegraphics[width = 50 mm]{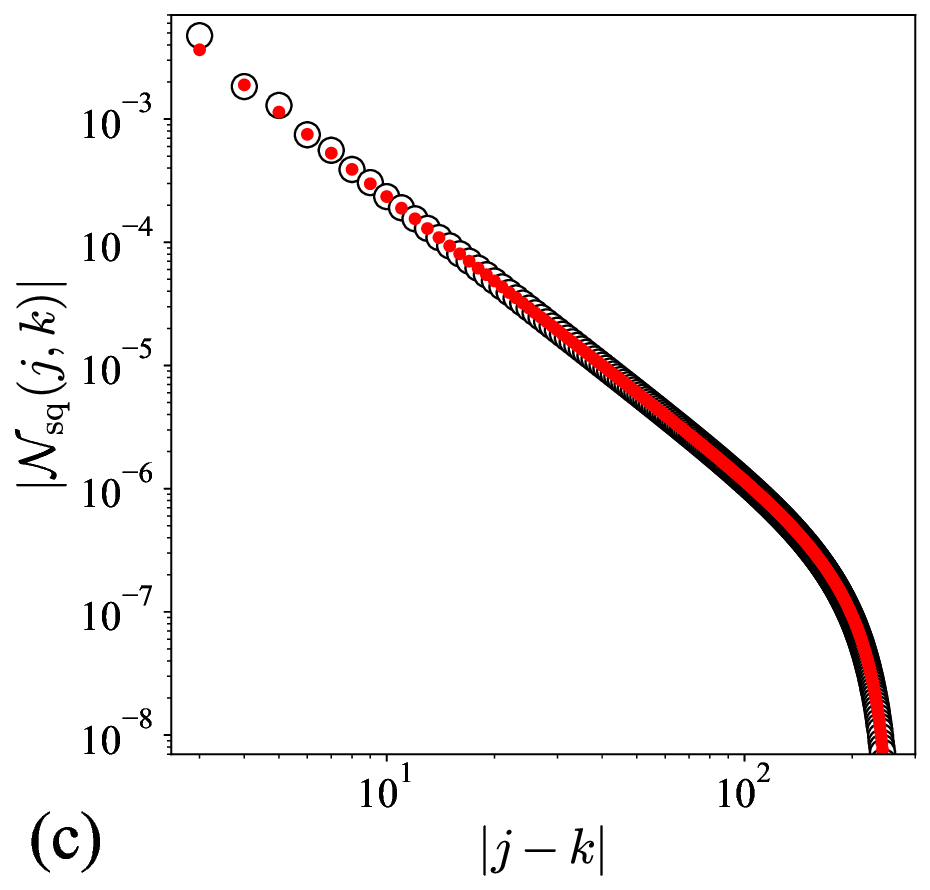}
\caption{
Two-spin correlation functions of the dimer and squared-spin operators, (a) $\langle \mathcal{N}_{\rm d}^z(j,k) \rangle$, (b) $\langle \mathcal{N}_{\rm d}^{xy}(j,k) \rangle$, and (c) $\langle \mathcal{N}_{\rm sq}(j,k) \rangle$ for $\Delta=1.50$ and $L=256$.
The absolute values are plotted as a function of $|j-k|$.
Open and solid circles respectively represent the DMRG data and their fits to the analytic forms Eqs.\ (\ref{eq:form_dimercorr}) and (\ref{eq:form_Sz2corr}).
}
\label{fig:Od_fits}
\end{center}
\end{figure*}

Finally, we present in Fig.\ \ref{fig:Od_fits} the two-point correlation functions of the dimer operators and the squared spin operator, $\mathcal{N}^z_{\rm d}(j,k)$, $\mathcal{N}^{xy}_{\rm d}(j,k)$, $\mathcal{N}_{\rm sq}(j,k)$, and their fits to Eqs.\ (\ref{eq:form_dimercorr}) and (\ref{eq:form_Sz2corr}) for $\Delta=1.50$ and $L=256$.
In the fitting, the exponent $K_+$ and the amplitude $A_{\rm d}^\mu$ or $A_2$ are taken as fitting parameters.
For $\mathcal{N}^z_{\rm d}(j,k)$ and $\mathcal{N}_{\rm sq}(j,k)$, the fitting results reproduce the numerical data very well, supporting the validity of the analytic formulas.
On the other hand, the fitting of $\mathcal{N}^{xy}_{\rm d}(j,k)$ seems to work poorly, and we have indeed found that the estimate of $K_+$ varies depending on the range of the data used for the fitting.
This poor fitting may be attributed to the small magnitude of $\mathcal{N}^{xy}_{\rm d}(j,k)$, by about two orders, compared to $\mathcal{N}^z_{\rm d}(j,k)$ and $\mathcal{N}_{\rm sq}(j,k)$, which makes the effect of short-range oscillating components relatively large.
%{\bf (**** add a dicsussion on the reason of the poor fitting of $\mathcal{O}^{xy}_{\rm d}(j)$, if needed and/or possible ****)}

\subsection{TLL parameter and correlation length}\label{subsec:exponent}

\begin{figure}
\begin{center}
\includegraphics[width = 73 mm]{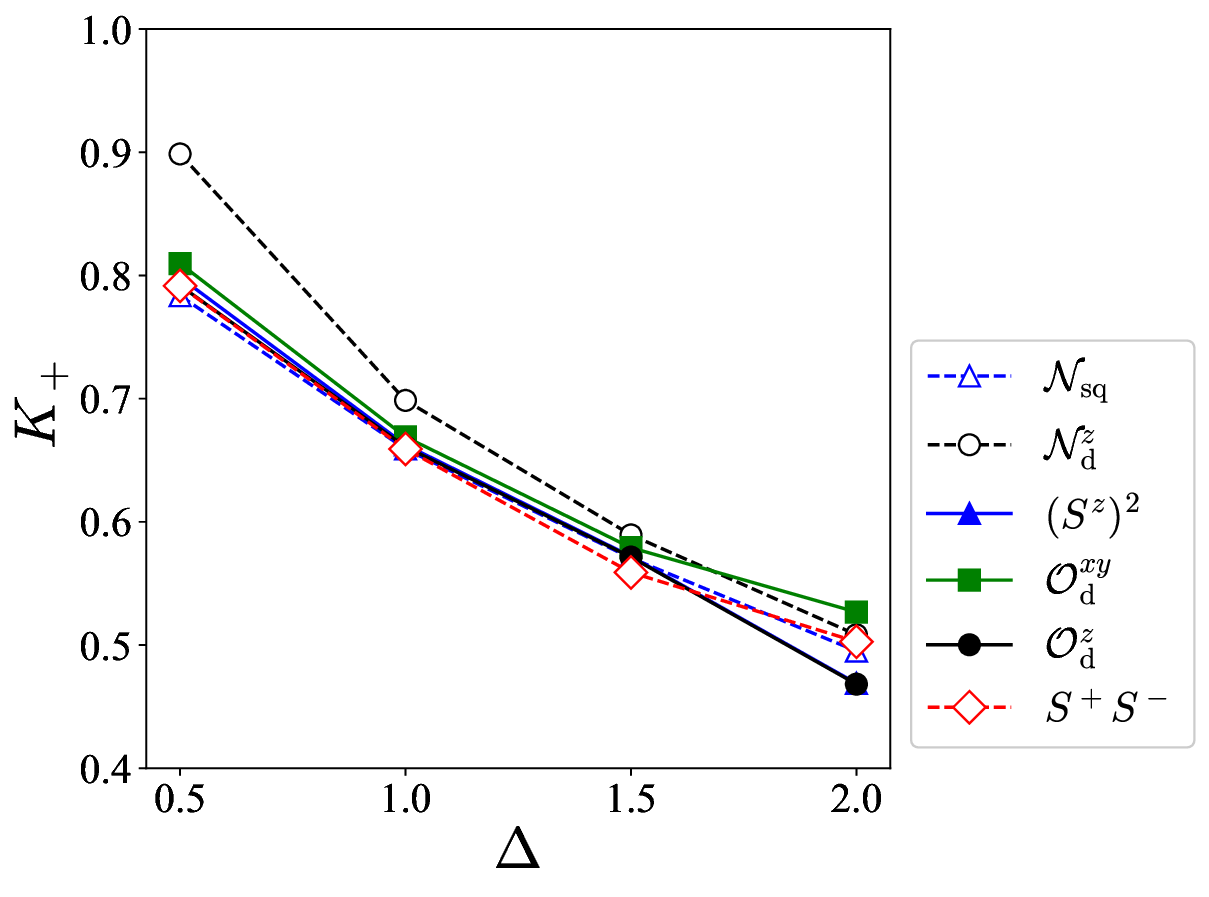}
\caption{
TLL parameter $K_+$ estimated by the fitting of the one-point and two-point correlation functions for $L=256$ as a function of $\Delta$.
}
\label{fig:K+}
\end{center}
\end{figure}

Figure\ \ref{fig:K+} shows the TLL parameter $K_+$ for $0.50 \le \Delta \le 2.00$ obtained from the fitting analysis of the one- and two-point correlation functions in the previous sections.
We note that the transverse spin operator $S^+_j$ [whose two-point correlation function has the decay exponent $1/(4K_+)$] and the dimer and squared-spin operators $\mathcal{O}^z_{\rm d}(j)$, $\mathcal{O}^{xy}_{\rm d}(j)$, and $(S^z_j)^2$ (the decay exponent $4K_+$) 
are dual to each other in the sense that their bosonized formulas
contain dual operators $\exp(\sqrt{\pi/2}\,\theta_+)$ and $\cos(\sqrt{8\pi}\,\phi_+)$ in the effective theory.
As seen in the figure, the estimates of $K_+$ from the transverse spin correlation function and the one-point functions of the dimer and squared-spin operators, for which the fitting of high quality was achieved, agree well with each other for $0.50 \le \Delta \le 1.50$.
The value of $K_+$ obtained from the two-point correlation function of $(S^z_j)^2$ is also in good agreement.
We also note that the correlation function of the longitudinal spin operator $S^z_j$ was reproduced well in Fig.~\ref{fig:SzSz_fits} by Eq.\ (\ref{eq:form_SzSzcor}) with $K_+$ estimated from the one-point function of $\langle \mathcal{O}^z_{\rm d}(j)\rangle$.
However, as $\Delta$ increases to $2.00$, the estimates of $K_+$ begin to deviate from each other.
This deviation may be qualitatively understood as the effect of the slow renormalization-group flows near the multicritical point, which makes the fitting of the data in finite-size chains less reliable.
%(See also the Appendix\ \ref{app:Delta-dep} for the fitting results for large $\Delta$.)

%{\bf (**** add the discussion on the deviation of $K_+$ from the correlation of $\mathcal{O}^z_{\rm d}(j)$, if possible ****)}

\begin{figure}
\begin{center}
\includegraphics[width = 73 mm]{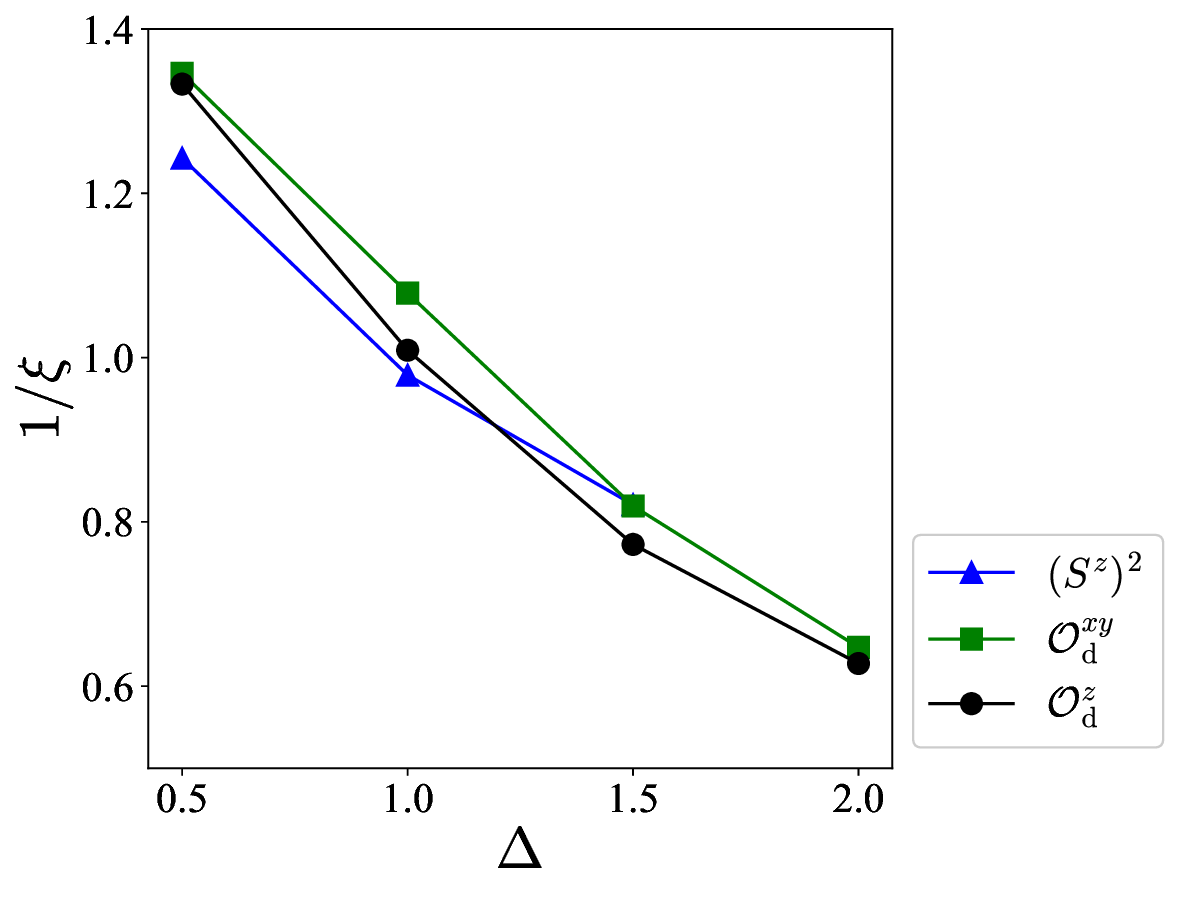}
\caption{
Inverse correlation length $1/\xi$ estimated from the exponentially-decaying boundary components in the one-point functions for $L=256$ as a function of $\Delta$.
}
\label{fig:1ovxi}
\end{center}
\end{figure}

We now discuss the correlation length that controls the decay of the boundary oscillating term of the one-point functions of the dimer operators $\mathcal{O}_{\rm d}^\mu(j)$ ($\mu=z,xy$) and the squared spin operators $(S^z_j)^2$.
We have determined the correlation length by fitting the data of $\langle \mathcal{O}_{\rm d}^\mu(j)\rangle_{\rm exp}$ [Eq.\ (\ref{eq:dimeroperator_exp})] and $\langle (S^z_j)^2\rangle_{\rm exp}$ [Eq.\ (\ref{eq:Sz2operator_exp})] near the boundary $j=1$ to the exponentially-decaying form \cite{fit_exp},
\begin{eqnarray}
\langle \mathcal{O}(j) \rangle_{\rm exp} = A_{\rm exp} (-1)^j \exp(-x_j/\xi),
\end{eqnarray}
where $x_j$ is the center position of the operators, i.e., $x_j=j+\frac{1}{2}$ for $\mathcal{O}(j)=\mathcal{O}_{\rm d}^\mu(j)$ and $x_j=j$ for $\mathcal{O}(j)=(S^z_j)^2$.
See Fig.\ \ref{fig:1p-c0-c1-c2} for the case of $\Delta=1.50$.
Figure\ \ref{fig:1ovxi} shows the estimates of the inverse correlation length obtained for $L=256$ as a function of $\Delta$.
We find that all one-point functions $\mathcal{O}_{\rm d}^\mu(j)$ ($\mu=z,xy$) and $(S^z_j)^2$ yield almost the same value of $1/\xi$, suggesting that the exponential decay of the boundary staggered term of those one-point functions stems from the same physics, presumably the excitation gap in the antisymmetric sector of the bosonic fields, $(\phi_-, \theta_-)$.
It is also reasonable that the correlation length increases as $\Delta$ increases and approaches the value of the multicritical point.
However, the fitting was not possible for $\langle (S^z_j)^2\rangle_{\rm exp}$ at $\Delta=2.0$ and for all the one-point functions at $\Delta=2.5$ since those $\langle \mathcal{O}(j)\rangle_{\rm exp}$ were not staggered even in the vicinity of the boundary.
This may be attributed to the insufficient precision in the fitting of the one-point functions $\langle \mathcal{O}(j)\rangle$ to Eqs.\ (\ref{eq:form_dimeroperator}) and (\ref{eq:form_Sz2operator}), which prevent us from removing the uniform components accurately in Eqs.\ (\ref{eq:dimeroperator_exp}) and (\ref{eq:Sz2operator_exp}).

%Besed on the results above, we conclude that the asymptotic properties of the model (\ref{eq:Ham}) at the critical point between the Haldane and large-$D$ phases are described correctly and consistently by the low-energy effective theory derived in Sec.\ \ref{sec:effective_theory}.

\subsection{Case with bond alternation}\label{subsec:Num_bondalternation}

To investigate the effect of bond alternation, we performed the DMRG calculation for the model (\ref{eq:Ham}) with the bond alternation term (\ref{H_delta}).
We treated the finite chains defined at the sites $1 \le j \le L$ under the open boundary conditions.
Note that the strength of the exchange interaction at the edge bonds is $1 + \delta$.
The calculations were carried out at the transition points between the Haldane and large-$D$ phases for $\Delta=1.50$ and several values of $\delta$ in the range $-0.100 \le \delta \le 0.100$.
The critical values $D_c$ for these parameters were obtained in Appendix\ \ref{app:crt_point}.
Below, we focus on the longitudinal two-spin correlation function $\langle S^z_j S^z_k \rangle$, in which the effect of bond alternation is particularly pronounced.
As Eq.\ (\ref{SzSz_thermodynamic_BA}) of $\langle S^z_j S^z_k \rangle$ in the thermodynamic limit indicates,
the bond alternation should induce the staggered component that decays algebraically with the exponent $4K_+$,
which is expected to be comparable to the exponent ($=2$) of the uniform component when $\Delta$ is larger than $1.50$ (see Fig.~\ref{fig:K+}).

Using the DMRG, we computed $\langle S^z_j S^z_k \rangle$ for the two sites $(j,k)$ satisfying $j+k=L$ or $L+1$, in the finite open chains with up to $L=256$ spins.
We then fit the DMRG data to the analytic form Eq.\ (\ref{eq:SzSz_open_form_BA}), taking $\tilde{a}$ and $K_+$ as fitting parameters.
For the fitting, we only use the data within the range $r_s \le |j-k| \le r_b$ for the three cases of $r_s = r_b = \frac{L}{8}, \frac{3L}{16}$, and $\frac{L}{4}$.
We take the averages of the values of $\tilde{a}$ and $K_+$ obtained from these three cases as the estimates of the two parameters.

\begin{figure}
\begin{center}
\includegraphics[width = 60 mm]{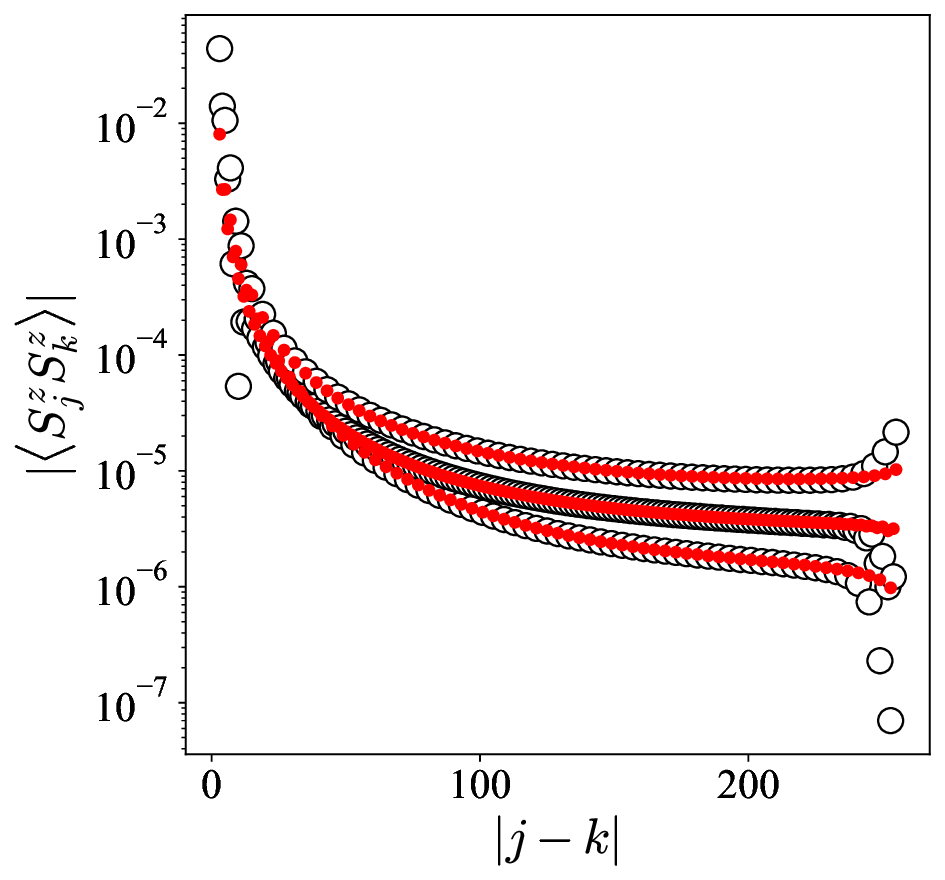}
\caption{
Longitudinal two-spin correlation function $\langle S^z_j S^z_k \rangle$ in the chain with bond alternation for $\Delta=1.50$, $\delta=0.050$, and $L=256$.
The absolute values are plotted as a function of $|j-k|$.
Open and solid circles respectively represent the DMRG data and their fits to the analytic form Eq.\ (\ref{eq:SzSz_open_form_BA}).
}
\label{fig:SzSzfit_BA}
\end{center}
\end{figure}

Figure\ \ref{fig:SzSzfit_BA} presents the DMRG data of $\langle S^z_j S^z_k \rangle$ and their fits to Eq.\ (\ref{eq:SzSz_open_form_BA}) at the critical point for $\Delta=1.50$ and $\delta=0.050$.
The DMRG data of $\langle S^z_j S^z_k \rangle$ with $j+k=L, L+1$, plotted as a function of $r=|j-k|$, clearly exhibits four-site periodic oscillations, in contrast to the $\delta=0$ case shown in Fig.~\ref{fig:SzSz_fits}.
This oscillating behavior is similar to that of the critical spin-$\frac12$ XXZ chain under the open boundary conditions \cite{HikiharaF1998} and can be explained as arising from the second and third terms with coefficients $\tilde{a}^2$ and $\tilde{a}$, respectively, in the analytic form Eq.\ (\ref{eq:SzSz_open_form_BA}).
Indeed, the DMRG data are reproduced very well by the fitting results over a wide range of $r=|j-k|$, indicating the validity of Eq.\ (\ref{eq:SzSz_open_form_BA}).
%We also emphasize that in the case of the uniform chain ($\delta=0$), such an oscillating behavior of four-site period is absent, as shown in Fig.\ \ref{fig:SzSz_fits}.

\begin{figure}
\begin{center}
\includegraphics[width = 60 mm]{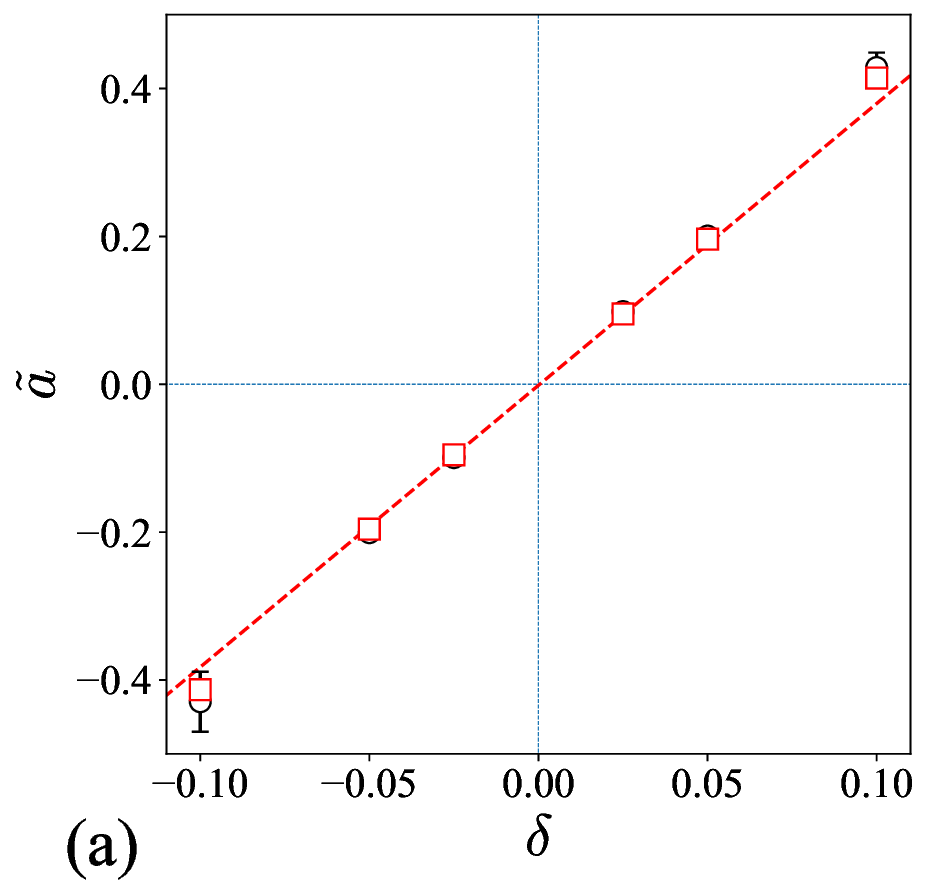}
\includegraphics[width = 60 mm]{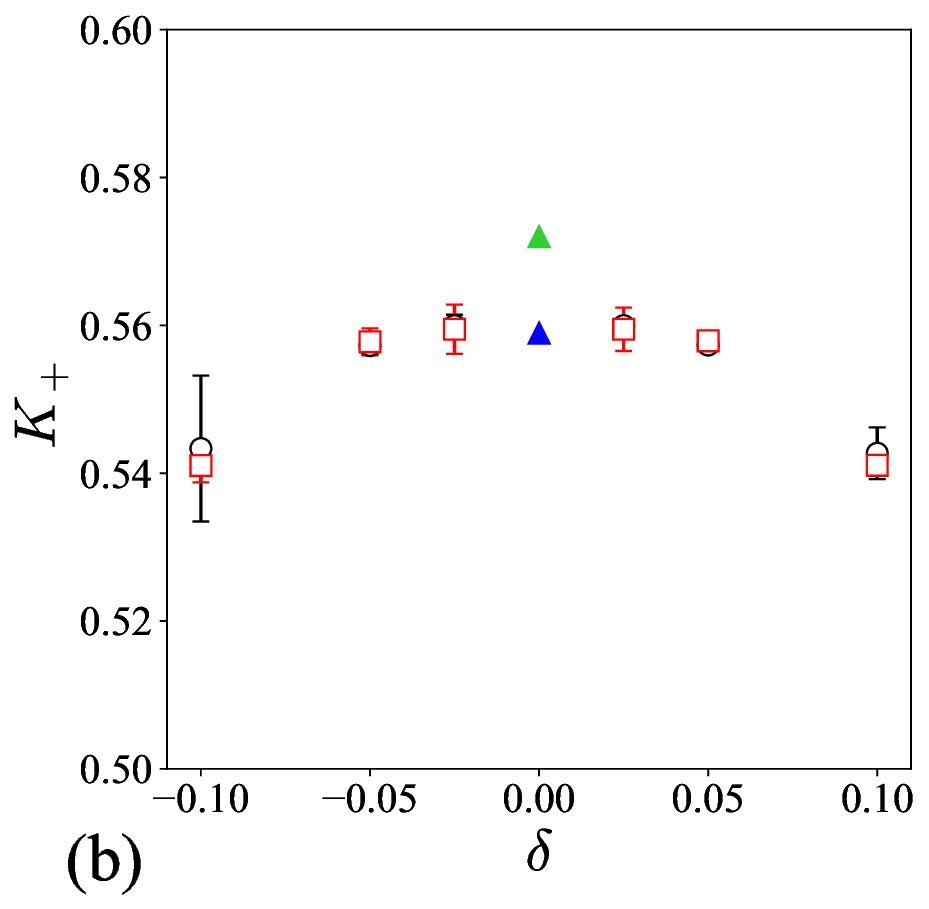}
\caption{
(a) Coefficient $\tilde{a}$ and (b) TLL parameter $K_+$ for $\Delta=1.50$ as a function of $\delta$.
Squares (red) and circles (black) represent the estimates obtained by the fitting of the data for $L=256$ and $L=128$, respectively.
Error bars represent the maximum deviation of the estimates of three cases of fitting from their average.
In (a), a red dotted line passing through the two data points at $\delta=-0.025$ and $0.025$ is drawn as a guide for the eye.
In (b), the blue and green triangles at $\delta=0$ respectively represent the estimates of $K_+$ from the fitting of $\langle S^+_j S^-_k \rangle$ and $(S^z_j)^2$ (the same data as those in Fig.\ \ref{fig:K+}).
}
\label{fig:aK_BA}
\end{center}
\end{figure}

Figure\ \ref{fig:aK_BA} shows the $\delta$ dependences of the estimates of $\tilde{a}$ and $K_+$ obtained from the fitting.
We find that $\tilde{a}$ exhibits a linear dependence on $\delta$, as predicted in Sec.\ \ref{subsec:bondalternation}.
The estimate of $K_+$ shows a slight $\delta$ dependence, which is symmetric with respect to $\delta=0$, as expected.

Based on the results presented above, we conclude that the effective theory presented in Sec.\ \ref{subsec:bondalternation} for the case with bond alternation has been numerically verified.

\section{Conclusion}\label{sec:conc}

In summary, we have investigated the quantum critical state realized at the phase transition between the Haldane and large-$D$ phases in the $S=1$ XXZ chain with uniaxial single-ion anisotropy.
Following the bosonization approach of Schulz\cite{Schulz1986}, we have described the low-energy effective theory for the critical state, which is the $c=1$ Gaussian theory, and derived analytic expressions for various correlation functions.
It is shown that the correlation functions at the quantum criticality exhibit the following peculiar properties.
In the correlation functions of the longitudinal spin operator $S^z_j$, the squared-spin operator $(S^z_j)^2$, and the dimer operators $\mathcal{O}^\mu_{\rm d}(j)$ ($\mu=z,xy$), only the uniform components decay algebraically, and the oscillating terms $\propto (-1)^r$ decay exponentially.
In contrast, the transverse two-spin correlation function $\langle S^+_j S^-_k \rangle$ exhibits an algebraic decaying behavior only in the staggered components, and its uniform components decay exponentially.
Then, we have performed numerical calculations using the DMRG method and shown that the correlation functions obtained numerically are fitted well by the analytic forms obtained using the bosonization approach.
We have thereby demonstrated the validity of the effective theory and determined the TLL parameter and the correlation length quantitatively.
The effect of the bond alternation on the quantum critical state at the transition has also been clarified.

The results of this paper are expected to be directly applicable to the spin-1/2 two-leg ladder with anisotropic rung couplings, which exhibits the same phase transition between the Haldane and large-$D$ phases in its phase diagram\cite{TonegawaOHS2017}.
Furthermore, a physics similar to that described in this paper may be realized in 1D spin models with multiple modes, where one gapless mode forms the Gaussian model and the other modes are gapped. 
For example, it has been shown that the spin-1/2 two-leg ladder \cite{HikiharaF2001} and $S=1$ chain \cite{Sato2006} under a magnetic field exhibit characteristic behaviors similar to our results; 
namely, some components of correlation functions exhibit an exponential decay due to a strongly-fluctuating boson field associated with the gapped mode, while only the components that are free from such fluctuating fields decay algebraically.
It would be interesting to explore whether or not peculiar properties by a similar mechanism appear in other 1D models with $S \ge 1$ spins or multi-leg ladder models.

From an experimental perspective, it has been established that the $S=1$ spin chain with single-ion anisotropy is realized in Ni(C$_2$H$_8$N$_2$)$_2$NO$_2$(ClO$_4$) (NENP)\cite{Renard1987}, Ni(C$_5$H$_{14}$N$_2$)$_2$N$_3$(PF$_6$) (NDMAP)\cite{HondaKNH2001}, and other several materials. (See, {\it e.g.}, \cite{YamashitaIM2000,WierschemS2014} and references therein.)
The realization of the $S=1$ spin chain has also been pursued in cold atom systems\cite{Sompet2022,Brechtelsbauer2025,Mogerle2025}.
The peculiar behaviors of correlation functions we have found in this study, such as the absence of algebraically-decaying terms in the staggered (uniform) components of the longitudinal (transverse) two-spin correlation function, can be detected through observables such as the dynamical structure factor measured by neutron scattering experiments.
We hope the results of the present work will stimulate further experimental studies.

\begin{acknowledgments}
The authors thank Kiyomi Okamoto for useful comments.
This work is partially supported by KAKENHI Grant Number JP24K06881.
The work of A. F. was supported in part by JST CREST (Grant No.\ JPMJCR19T2).
\end{acknowledgments}

\appendix
\section{Derivation of effective model (\ref{eq:sum_H+_H-_Hpm})}\label{app:bosonization_details}

In this appendix, we present details of the calculation to derive the bosonized effective Hamiltonian (\ref{eq:sum_H+_H-_Hpm}) with Eq.\ (\ref{eq:boson_H+_H-_Hpm}).
To that end, we first represent the $S=1$ spin operator $\bm{S}_j$ by a sum of two $S=1/2$ spin operators $\bm{s}_{n,j}$ ($n=1,2$) as in Eq.\ (\ref{S=s1+s2}).
Substituting Eq.\ (\ref{S=s1+s2}) into Eq.\ (\ref{eq:Ham}), we obtain the Hamiltonian in terms of $\bm{s}_{n,j}$ as
\begin{equation}
H=\sum_{j}\left[h_1(j)+h_2(j)+h_{12}(j)+h_D(j)+\frac{D}{2}\right],
\end{equation}
where
\begin{subequations}
\begin{align}
h_n(j)
%&=s_{n,j}^x s_{n,j+1}^x + s_{n,j}^y s_{n,j}^y + \Delta s_{n,j}^zs_{n,j+1}^z
%\nonumber\\
&=
\frac12(s_{n,j}^+s_{n,j+1}^- + s_{n,j}^-s_{n,j+1}^+)+\Delta s_{n,j}^zs_{n,j+1}^z,
%\quad(n=1,2),
\label{H_n}
\\
h_{12}(j)&=
\frac12(
s_{1,j}^+s_{2,j+1}^- +s_{1,j}^-s_{2,j+1}^+ 
\nonumber \\
&\qquad
+s_{2,j}^+s_{1,j+1}^- +s_{2,j}^-s_{1,j+1}^+)
\nonumber \\
&~~~+\Delta(s_{1,j}^zs_{2,j+1}^z+s_{2,j}^zs_{1,j+1}^z),
\label{H_12}
\\
h_D(j)&=2Ds_{1,j}^zs_{2,j}^z.
\label{H_D}
\end{align}
\end{subequations}
Here we have defined
% $S_j^\pm = S_j^x \pm i S_j^y$ and 
$s_{n,j}^\pm = s_{n,j}^x \pm i s_{n,j}^y$.

The exchange interactions $h_1(j)$ and $h_2(j)$ in Eq.~(\ref{H_n}) are the Hamiltonian densities of the two independent spin-$\frac12$ XXZ chains, which can be readily bosonized.
Using the notation in Eq.\ (\ref{eq:s_n,j}), the bosonic-field representations of $h_n(j)$ in lowest order in $\Delta$ are obtained as
\begin{align}
h_n(j)=&\,\frac{1}{2}\left[
\left(1+\frac{4\Delta}{\pi}\right)\!\left(\frac{d\phi_n}{dx}\right)^2
+\left(\frac{d\theta_n}{dx}\right)^2\right]
\nonumber \\
&+\frac{a^2\Delta}{2}\cos(\sqrt{16\pi}\phi_n)
\nonumber \\
&+(-1)^jd(2+\Delta)\cos(\sqrt{4\pi}\phi_n),
\label{h_n(j)}
\end{align}
where 
%$\alpha$ is a short-distance cutoff of the order of the lattice spacing, and 
$d$ is a real parameter related to the dimer correlation; see e.g., Ref.~\cite{HikiharaFL2017,MudryFMH2019}.
We have kept the last oscillating term for later discussions.
Furthermore, by introducing the linear combinations of the bosonic fields $\phi_\pm(x)$ and $\theta_\pm(x)$ as in Eq.\ (\ref{eq:phi_theta_pm}), we rewrite the Hamiltonian density of the two spin-$\frac{1}{2}$ chains as
\begin{align}
\sum_{n=1,2}\! h_n(j)
%+h_2(j)
%\nonumber \\
%&&~~~~
=&{}
\frac{1}{2}\!\left\{
\left(1+\frac{4\Delta}{\pi}\right)\!\left[
\left(\frac{d\phi_+}{dx}\right)^2+\left(\frac{d\phi_-}{dx}\right)^2
\,\right] \right.
\nonumber\\
&~~~~~~~~~~~~~ \left.
+\left(\frac{d\theta_+}{dx}\right)^2+\left(\frac{d\theta_-}{dx}\right)^2
\right\}
\nonumber\\
&
+a^2 \Delta
 \cos(\sqrt{8\pi}\phi_+)\cos(\sqrt{8\pi}\phi_-)
\nonumber\\
&
+2d(-1)^j(2+\Delta)
 \cos(\sqrt{2\pi}\phi_+)\cos(\sqrt{2\pi}\phi_-).
\nonumber \\
\label{h_1+h_2}
\end{align}

Next, we evaluate the exchange interaction between the two spin-$\frac12$ chains, $h_{12}(j)$.
The transverse exchange interaction in $h_{12}(j)$ is bosonized as
\begin{align}
\epsilon_{xy}=&\,
\frac12(
s_{1,j}^+s_{2,j+1}^- + s_{1,j}^-s_{2,j+1}^+ +
s_{2,j}^+s_{1,j+1}^- + s_{2,j}^-s_{1,j+1}^+
)
\nonumber\\
%&=& -\frac12
%\left(e^{i\sqrt{\pi}(\theta_1-\theta_2)}+e^{-i\sqrt{\pi}(\theta_1-\theta_2)}\right)
%\nonumber \\
%&&~~~~~\times \left[b^2+\tilde{b}^2\sin(\sqrt{4\pi}\phi_1)\sin(\sqrt{4\pi}\phi_2)
%\right]
%\nonumber\\
=& -2b^2\cos(\sqrt{2\pi}\theta_-)
\nonumber \\
&
-\tilde{b}^2\cos(\sqrt{2\pi}\theta_-)\left[\cos(\sqrt{8\pi}\phi_-)-\cos(\sqrt{8\pi}\phi_+)\right].
\nonumber\\&
\label{epsilon_xy}
\end{align}
Here, the oscillating terms $\propto(-1)^jb\tilde{b}$ cancel out under the approximation
$\phi_n(j+1)\approx\phi_n(j)$ and $\theta_n(j+1)\approx\theta_n(j)$. 
With the same approximation, the longitudinal exchange interaction becomes
\begin{align}
\epsilon_z=&\,
s_{1,j}^zs_{2,j+1}^z+s_{2,j}^zs_{1,j+1}^z
\nonumber\\
%&=&
%\frac{2}{\pi}\frac{d\phi_1}{dx}\frac{d\phi_2}{dx}
%-2a^2\sin(\sqrt{4\pi}\phi_1)\sin(\sqrt{4\pi}\phi_2)
%\nonumber\\
=&\,
\frac{1}{\pi}\left[
\left(\frac{d\phi_+}{dx}\right)^2-\left(\frac{d\phi_-}{dx}\right)^2
\right]
\nonumber \\
&+a^2\left[\cos(\sqrt{8\pi}\phi_+)-\cos(\sqrt{8\pi}\phi_-)\right],
\label{epsilon_z}
\end{align}
where the oscillating terms $\propto a(-1)^j$ cancel again using
the approximation $\phi_n(j+1)\approx\phi_n(j)$.
Similarly, the single-ion anisotropy interaction is bosonized as
\begin{align}
\frac{h_D}{D}
%&=&
%\frac{2}{\pi}\frac{d\phi_1}{dx}\frac{d\phi_2}{dx}
%+2a^2\sin(\sqrt{4\pi}\phi_1)\sin(\sqrt{4\pi}\phi_2)
%\nonumber \\
%&&~+\frac{2a(-1)^j}{\sqrt\pi}
%\left(
%\frac{d\phi_1}{dx}\sin(\sqrt{4\pi}\phi_2)+\frac{d\phi_2}{dx}\sin(\sqrt{4\pi}\phi_1)
%\right)
%\nonumber\\
=&\,
\frac{1}{\pi}\left[
\left(\frac{d\phi_+}{dx}\right)^2-\left(\frac{d\phi_-}{dx}\right)^2
\right]
\nonumber \\
&
-a^2\left[\cos(\sqrt{8\pi}\phi_+)-\cos(\sqrt{8\pi}\phi_-)\right]
\nonumber\\
&+a(-1)^j\sqrt{\frac{8}{\pi}}\left[
\frac{d\phi_+}{dx}\sin(\sqrt{2\pi}\phi_+)\cos(\sqrt{2\pi}\phi_-)
\right. \nonumber \\
&\qquad\qquad\qquad\left.
-\frac{d\phi_-}{dx}\cos(\sqrt{2\pi}\phi_+)\sin(\sqrt{2\pi}\phi_-)
\right].
\nonumber\\ &
\label{h_D}
\end{align}
From Eqs.\ (\ref{h_1+h_2}), (\ref{epsilon_xy}), (\ref{epsilon_z}), and (\ref{h_D}), we obtain the effective Hamiltonian (\ref{eq:sum_H+_H-_Hpm}) with Eq.\ (\ref{eq:boson_H+_H-_Hpm}).

\section{Derivation of correlation functions in finite open chains}\label{app:finite_open_chain}

Here, we summarize the calculation of the ground-state expection values of various operators in the model (\ref{eq:Ham}) with $L$ spins under open boundary conditions at the transition between the Haldane and large-$D$ phases.
To that end, we take $H_+^{(0)}$ in Eq.\ (\ref{H_+^0}) as the effective theory and impose the Dirichlet boundary conditions on the bosonic field $\phi_+(x)$
\cite{EggertA1992,HikiharaF1998,HikiharaF2001,HikiharaF2004},
\begin{eqnarray}
\phi_+(0)=\phi_+(L+1)=0,
\label{eq:Dirichlet_phi+}
\end{eqnarray}
which represents the absence of spins at $j=0$ and $j=L+1$ in the finite spin chain.
Under these boundary conditions,
one can expand the fields $\phi_+(x)$ and $\theta_+(x)$ as
\begin{subequations}
\label{eq:mode_expansion}
\begin{align}
&\phi_+(x)
= \frac{x}{L+1}\phi_0 + \sqrt{K_+} \sum_{n=1}^\infty e^{-\frac{\zeta n}{2}}
\frac{\sin(q_n x)}{\sqrt{\pi n}} (a_n + a_n^\dagger),
\label{eq:mode_expansion_phi+}\\
& \theta_+(x)
= \theta_0 + \frac{i}{\sqrt{K_+}} \sum_{n=1}^\infty e^{-\frac{\zeta n}{2}}
\frac{\cos(q_n x)}{\sqrt{\pi n}} (a_n - a_n^\dagger),
\label{eq:mode_expansion_theta+}
\end{align}
\end{subequations}
where $q_n=\pi n/(L+1)$, $[\theta_0, \phi_0] = i$, $[a_n, a_m^\dagger] = \delta_{nm}$,
and we have introduced a small positive constant $\zeta$ for regularization.
Substituting Eqs.\ (\ref{eq:mode_expansion_phi+}) and (\ref{eq:mode_expansion_theta+}) into Eq.\ (\ref{H_+^0}), one can get the expression of the effective Hamiltonian $H_+^{(0)}$ in terms of $a_n$ and $\phi_0$,
\begin{eqnarray}
\widetilde{H}_+^{(0)} = \sum_{n=1}^\infty v_+ q_n a_n^\dagger a_n
+ \frac{v_+ \phi_0^2}{2K_+ (L+1)} - \frac{\pi v_+}{24(L+1)}.
\end{eqnarray}
Thus, the ground state of the spin chain is defined as the null state of $a_n$ and $\phi_0$, i.e., $a_n |0\rangle = \phi_0|0\rangle = 0$.

The ground-state expectation values of correlation functions in the finite open chains are obtained by subtituting the mode expansion Eq.~(\ref{eq:mode_expansion}) into the bosonic representations of the corresponding operators presented in Sec.\ \ref{subsec:corr_thermo} and taking their average in the vacuum $|0\rangle$.
We refer the reader to Refs.\ \cite{HikiharaF2001,HikiharaF2004,HikiharaFL2017} for the details of the calculation.
The resultant formulas of one- and two-point correlation functions are given in Sec.\ \ref{subsec:corr_openchain}.

\section{Critical point}\label{app:crt_point}

\begin{figure}
\begin{center}
\includegraphics[width = 65 mm]{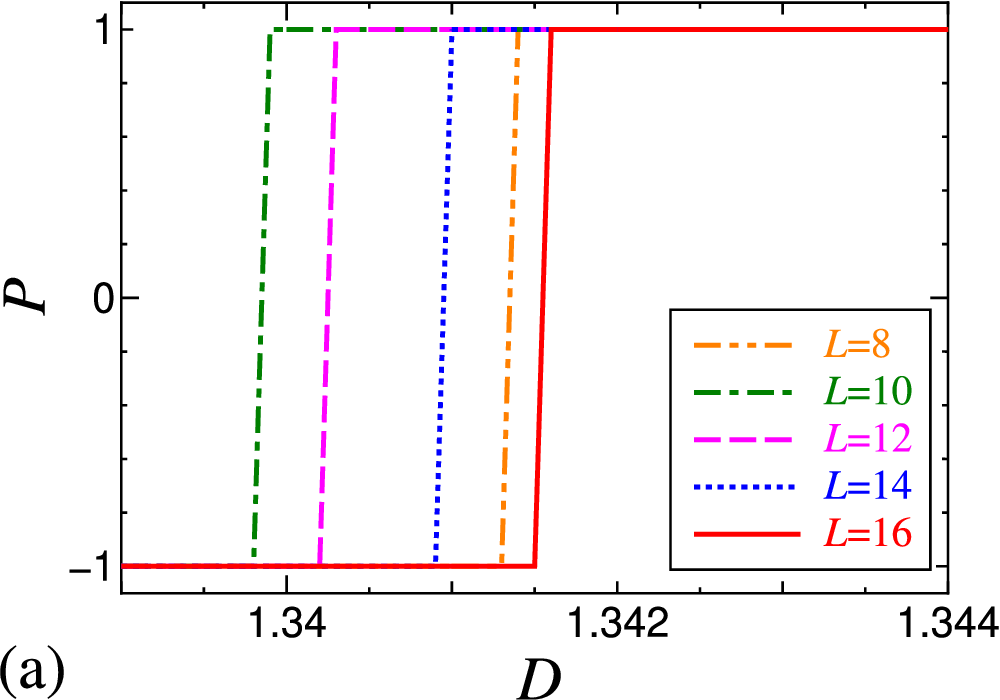}
\includegraphics[width = 65 mm]{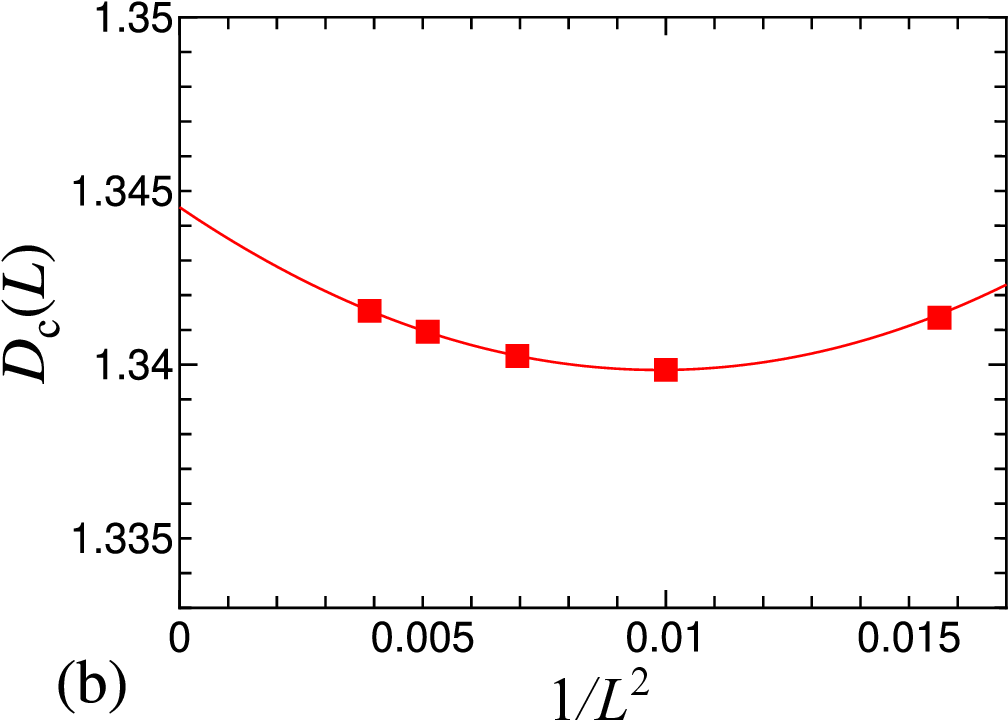}
\caption{
(a) Eigenvalue $P$ of the inversion operator $\hat{P}$ for $\Delta=1.50$ and $L=8$-$16$ as a function of $D$.
(b) Critical point $D_{\rm c}(L)$ for $\Delta=1.50$ as a function of $1/L^2$.
Solid curve represents the result of the fitting to the form $D_{\rm c}(L) = D_{\rm c} + e_1/L^2 + e_2/L^4$.
}
\label{fig:Dc}
\end{center}
\end{figure}

We determined the critical points between the Haldane and large-$D$ phases in the model (\ref{eq:Ham}) by the twisted-boundary
method \cite{KitazawaN1997a,KitazawaN1997b,ChenHS2000,ChenHS2008,ChenHS2003}.
The method is based on the fact that the ground state of the model (\ref{eq:Ham}) under the twisted boundary conditions $S^\pm_{L+1} = - S^\pm_1, S^z_{L+1} = S^z_1$ is an eigenstate of the inversion operater $\hat{P}$ with the eigenvalue $P=-1$ ($P=1$) for the Haldane (large-$D$) phase.
Here, the inversion operator $\hat{P}$ converts the site order as ${\bm S}_i \to {\bm S}_{L+1-i}$.
We used the Lanczos method to calculate the lowest-energy state of the model (\ref{eq:Ham}) under the twisted boundary conditions within the subspace of zero total magnetization, which is supposed to be the ground state.
The system size treated was up to $L=16$.
%Performing the calculation for the parameter lines with the fixed $\Delta$ with , 
We evaluated the eigenvalue $P$ as a function of $D$ for a fixed $\Delta$ while increasing $D$ in intervals of $0.0001$.
We then determined the critical point $D_{\rm c}(L)$ for the $L$-site system as the point where the eigenvalue $P$ jumps from $P=-1$ to $P=1$.
Finally, we extrapolated the critical values $D_{\rm c}(L)$ by fitting the data for $L=10, 12, 14,$ and $16$ to the form $D_{\rm c}(L) = D_{\rm c} + e_1/L^2 + e_2/L^4$.
Figure\ \ref{fig:Dc} shows the $D$ dependence of the eigenvalue $P$ and the extrapolation of $D_{\rm c}(L)$ for $\Delta=1.50$.
The extrapolated values $D_{\rm c}$ as a function of $\Delta$ are presented in Table\ \ref{tab:Dc}.
We use those values as the model parameters $(\Delta, D_{\rm c}(\Delta))$ at the critical points between the Haldane and large-$D$ phases.

\begin{figure}
\begin{center}
\includegraphics[width = 60 mm]{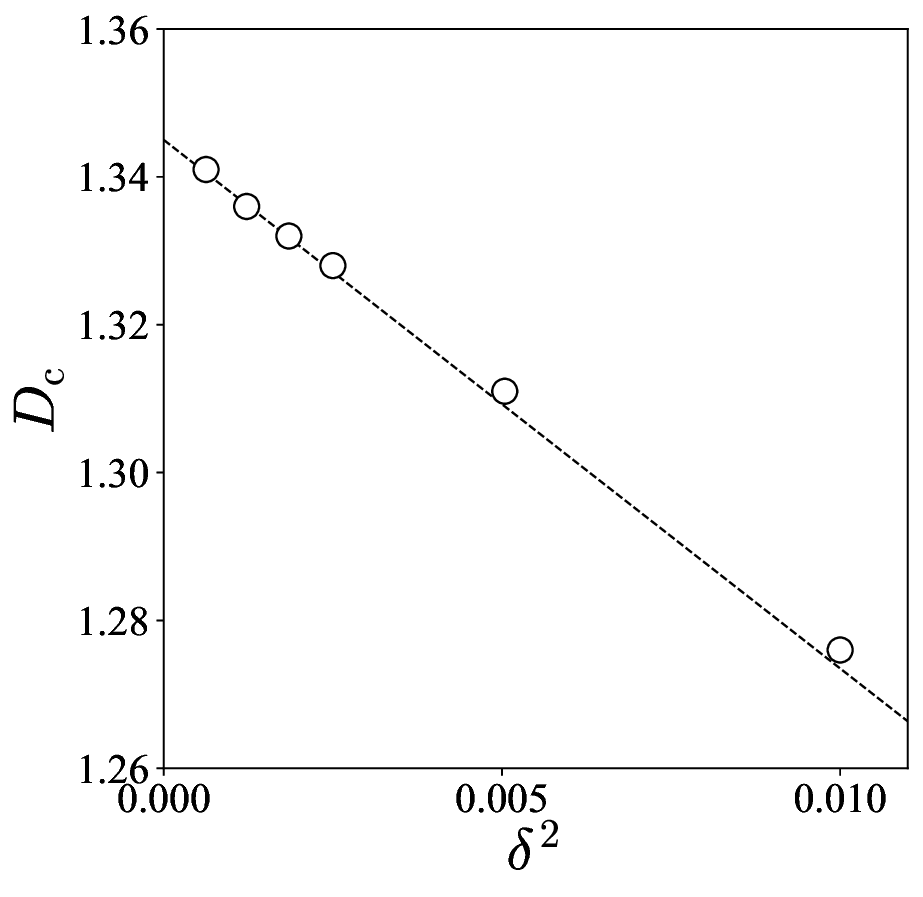}
\caption{
$D_{\rm c}(\Delta, \delta)$ for $\Delta=1.50$ and several values of $\delta^2$.
Dotted line represents the fitting result of the data for two smallest $\delta$'s to the form $D_{\rm c}(1,50, \delta) - D_{\rm c}(1.50, 0) = C_\delta \delta^2$ with $D_{\rm c}(1.50, 0) = 1.345$.
}
\label{fig:Dc_delta}
\end{center}
\end{figure}

We also used the same method as above to determine the critical points
between the Haldane and large-$D$ phases for the model (\ref{eq:Ham}) with the bond-alternation term (\ref{H_delta}): $H + H_{\delta}$. 
The critical value $D_c(\Delta,\delta)$ was determined 
% for the parameter lines specified by $(\Delta, \delta)$ 
for $\Delta=1.50$ and several values of $\delta$.
For this analysis, there were two ways to impose the twisted boundary condition: (i) on a bond with strength $1+\delta$ and (ii) on a bond with strength $1-\delta$.
We tried both and confirmed that the critical values $D_c$ coincide.
Figure\ \ref{fig:Dc_delta} shows the critical value $D_c(\Delta,\delta)$ as a function of $\delta^2$ for $\Delta=1.50$.
We find that the shift of $D_c(\Delta=1.50,\delta)$ from the value at the uniform case, $D_c(\Delta=1.50,0)=1.345$, is proportional to $\delta^2$, as discussed in Sec.\ \ref{subsec:bondalternation}.
We use those values of $D_c(\Delta,\delta)$ for $\Delta=1.50$ and $\delta=0.025, 0.050, 0.100$ for the analysis in Sec.\ \ref{subsec:Num_bondalternation}.

\begin{table}
\caption{
The extrapolated values $D_{\rm c}$ of the critical point between the Haldane and large-$D$ phases for $\Delta=0.50, 1.00, \ldots, 2.50$.
}
\label{tab:Dc}
\begin{center}
\begin{tabular}{cc}
\hline
\hline
 $\Delta$ & $D_{\rm c}$ \\
\hline
0.50 & 0.635  \\
1.00 & 0.969  \\
1.50 & 1.345  \\
2.00 & 1.757  \\
2.50 & 2.204  \\
\hline
\hline
\end{tabular}
\end{center}
\end{table}

%\bibliography{HLDcrt-arXiv_v2_Ref}

\begin{thebibliography}{41}
\bibitem [{fit()}]{fit_exp}%
  \BibitemOpen
  \bibinfo {note} {For the fitting, we basically used the data of $\langle
  \mathcal{O}(j)\rangle_{\rm exp}$ for $1 \le j \le 8$. However, for
  $\Delta=0.50$, the correlation length was so short that $\langle
  \mathcal{O}(j)\rangle_{\rm exp}$ for $\mathcal{O}(j)=\mathcal{O}_{\rm
  d}^{xy}(j)$ and $(S^z_j)^2$ exhibited non-staggered behavior at a large $j$.
  Accordingly, in the fitting of those operators for $\Delta=0.50$, we used the
  data of $\langle \mathcal{O}_{\rm d}^{xy}(j) \rangle_{\rm exp}$ for $1 \le j
  \le 7$ and $\langle (S^z_j)^2 \rangle_{\rm exp}$ for $1 \le j \le
  6$.}\BibitemShut {Stop}%
\end{thebibliography}

\end{document}